\documentclass[a4,11pt]{article}
\pdfoutput=1 
\usepackage{jcappub}
\usepackage{graphicx}
\usepackage{color, graphics}
\usepackage{amsmath,amssymb,amsfonts}
\usepackage{acronym}
\usepackage{mathrsfs}
 \usepackage{hyperref}
 \usepackage{units}
\usepackage{enumerate}   
\usepackage{etoolbox}

\def\beq{\begin{equation}\begin{aligned}}
\def\eeq{\end{aligned}\end{equation}}

\makeatletter
\renewcommand\Huge{\@setfontsize\Huge{24pt}{23pt}} 
\makeatother

\begin{document}

\title{\scalebox{0.97}{Constraining the Coexistence of Freeze-in}\\ \scalebox{0.97}{Dark Matter and Primordial Black Holes}}
\author[1]{Prolay Chanda,}
\affiliation[1]{Tata Institute of Fundamental Research, Homi Bhabha Road, Mumbai 400005, India}
\author[2]{Sagnik Mukherjee}
\author[2]{and James Unwin}
\affiliation[2]{Department of Physics,  University of Illinois Chicago, Chicago, IL 60607, USA}

\abstract{Particle dark matter and primordial black holes (PBH) might coexist with appreciable cosmic abundances, with both contributing to the 
observed dark matter density $\Omega_{\rm DM}$. Large populations of PBH (with $\Omega_{\rm PBH}\sim \Omega_{\rm DM}$) are tightly constrained for PBH heavier than $10^{-11} M_\odot$. However, large fractional abundances with  $ f_{\rm PBH}\simeq \Omega_{\rm PBH}/\Omega_{\rm DM}\sim0.01$ are consistent with the limits on PBH for a wide range of PBH masses. Scenarios with significant populations of both particle dark matter and PBH are intriguing. Notably, if the particle dark matter has interactions with the Standard Model, new constraints arise due to pair-annihilations that are enhanced by the PBHs, resulting in dark matter indirect detection constraints on $f_{\rm PBH}$. Here we derive the bounds on mixed scenarios in which PBHs coexist with particle dark matter whose relic abundance is set via freeze-in  (``FIMPs''). We show that while the restrictions on $f_{\rm PBH}$  are less constraining for FIMPs than WIMPs, modest bounds still arise for large classes of models. We examine both IR and UV freeze-in scenarios, including the case of ``superheavy'' particle dark matter with PeV scale mass.
}

\maketitle

\newpage
\section{Introduction}

The dark matter abundance $\Omega_{\rm DM}h^2$ inferred in astrophysical and cosmological observations could be due to more than a single source. One interesting possibility is that whilst particle dark matter could provide the bulk of $\Omega_{\rm DM}h^2$, a population of primordial black holes (PBH) \cite{Zeldovich:1967lct, Hawking:1971ei,Carr:1974nx,Carr:1975} might contribute a significant fraction, quantified in terms of the fractional abundance
\beq
f_{\rm PBH}\equiv\frac{\Omega_{\rm PBH}}{\Omega_{\rm DM}}\equiv\frac{\Omega_{\rm PBH}}{\Omega_{\rm PBH}+\Omega_{\rm pDM}}~,
\eeq
where $\Omega_{\rm DM}h^2\approx0.12$ \cite{Planck:2018vyg} is the dark matter density inferred from cosmology and  $\Omega_{\rm pDM}$ is the contribution from particle dark matter. In what follows, we assume  $\Omega_{\rm PBH}\ll\Omega_{\rm pDM}\approx\Omega_{\rm DM}$. Assuming that the mass spectrum of PBH masses is monochromatic, one typically requires $\Omega_{\rm PBH}\lesssim0.1$ to evade astrophysical constraints on PBH  \cite{Green:2024bam}, so this is a reasonable assumption for scenarios with both PBH and particle dark matter components.

Limits on scenarios with both PBH and particle dark matter have been well explored in the context of classic Weakly Interacting Massive Particle (WIMP) dark matter \cite{Mack:2006gz,Ricotti:2007jk,Ricotti:2009bs,Eroshenko:2016yve,Lacki:2010zf,Chanda:2022hls,Kadota:2021jhg,Adamek:2019gns,Carr:2020mqm,Gines:2022qzy,Tashiro:2021xnj,Boudaud:2021irr,Kadota:2022cij}. Notably, the general conclusion is that unless the fractional abundance of PBHs is diminutively small then this mixed scenario is strongly excluded from various astrophysical observations, in particular constraints from extragalactic gamma-ray. 
While the WIMP paradigm provides a compelling narrative to explain the observation of dark matter, it has increasingly come into tension with dedicated direct detection searches. Furthermore, there is a great variety of dark matter scenarios that provide alternative explanations for the origin and nature of dark matter particles.  
This paper presents the first dedicated study\footnote{Constraints on FIMPs due to annihilations around PBH were briefly examined in \cite{Scholtz:2019csj}, but strictly for the case of an Earth-mass PBH within the Solar System (in the context of the ``Planet 9'' hypothesis).} to apply indirect detection methods to constrain annihilations of non-WIMP dark matter around PBH, focusing on freeze-in dark matter \cite{Hall:2009bx}, or Feebly Interacting Massive Particles (FIMPs).

In freeze-in models, one assumes the initial abundance of dark matter particles is negligible. An appreciable abundance of dark matter particles is subsequently produced via interactions in the thermal bath. In order for this ``freeze-in'' process to entirely determine the late-time relic abundance it is important that the dark matter states do not undergo freeze-out, either through annihilations in the hidden sector or due to the dark matter particles reaching thermal equilibrium with the visible sector. Notably, in order to avoid sector equilibration, one requires that the interaction rate between the visible sector and hidden sector is highly suppressed, which typically implies that intra-sector couplings should be very small.
For minimal freeze-in scenarios, there are two general possibilities:
\begin{itemize}
\item IR freeze-in: The portal operator is renormalisble and the FIMP production rate is controlled by the coupling constants and masses of the various states involved.
\item UV freeze-in: The portal operator is non-renormalisble and the FIMP production rate is controlled by the mass scale suppressing the operator.
\end{itemize}
Notably, one should expect very different limits for FIMPs. While WIMPs have couplings $g\sim \mathcal{O}(1)$ couples such that are initially in equilibrium with the thermal bath, FIMPs are required to be always decoupled from the thermal bath, implying tiny (effective) couplings $\lambda\ll1$. This impacts both the cosmological history of the particle dark matter and the rate of observable signals, both of which will alter the constraints.

This paper is structured as follows: in Section \ref{S2}, we recap the formation of dark matter halos around PBH. We highlight deviations from standard treatments to occur due to the assumption that the relic density of dark matter particles is established via freeze-in. In Section \ref{S3}, we use these halo profiles to derive constraints on various FIMP models; we consider both renormalisable (IR) freeze-in and non-renormalisble UV freeze-in \cite{Hall:2009bx,Elahi:2014fsa}. These bounds are derived from extra-galactic gamma-ray observations due to Fermi-LAT  \cite{Fermi-LAT:2014ryh}. As well as deriving constraints on electroweak scale FIMPs, we also examine the limits on PeV scale particle dark matter. Section \ref{S4} provides some concluding remarks.

\section{Freeze-in and the PBH Dark Matter Halo}
\label{S2}

For both UV freeze-in and IR freeze-in the majority of FIMPs are produced at some characteristic temperature scale. In the case of UV freeze-in, this temperature scale is related to the highest temperature that the thermal bath attains after inflationary reheating. Since at lower temperatures, the non-renomormalisable portal operator becomes increasingly suppressed. 
In contrast, for IR freeze-in, the production mechanism typically switches off once the thermal bath cools below some mass scale, being either the mass of the FIMP $m_{\rm DM}$ or the mass of the mediator $m_M$ related to the renomormalisable portal interaction. For IR freeze-in, the majority of the FIMP population is produced immediately prior to the termination of the production process, at a temperature $T\sim {\rm max}(m_{\rm DM},m_M)$. 
When the dark matter particles are produced can have implications for the profile of the dark matter particle halo around PBH. In this paper, we shall assume that the freeze-in is instantaneous, at a characteristic time scale $t_{{\rm FI}}$. Thus prior to $t_{{\rm FI}}$ the FIMP abundance of the universe is negligible, and for $t>t_{{\rm FI}}$ the FIMP abundance accounts for the majority of the observed value of $\Omega_{\rm DM}$ (with the remainder being due to the PBH abundance). We assume that the FIMPs thermalise after production such that the hidden sector has a well-defined temperature $T'$.  To avoid sector equilibration, the hidden sector temperature $T'$ should always be much colder than the visible sector temperature $T$.

\subsection{Initial FIMP density distribution around PBHs}
\label{S2.1}

Let us first discuss the case of PBH halo formation in the traditional context of WIMP. Starting from a uniform distribution around a PBH, this density profile will subsequently evolve due to the gravitational pull of the central PBH. At first approximation, this causes the density of the halo to evolve from uniformity to a $\rho\propto r^{-9/4}$ power-law \cite{Bertschinger:1985pd} or ``spike'' profile.  A  more careful analysis shows that one obtains broken power laws. Moreover, one finds that there are three possibilities with distinct broken power-laws, with the form depending on the mass of the PBH and properties of the particle dark matter \cite{Boudaud:2021irr}.   We summarise the three cases in Appendix \ref{ApA}, recasting them in terms of FIMPs rather than WIMPs.

 The first point of distinction between FIMPs and WIMPs comes from the role that kinetic decoupling plays in the formation of WIMP halos around PBHs. Initially, the WIMPs are relativistic; as such, they free-stream. At the point of WIMP kinetic decoupling $t_{\rm kd}$, PBH will capture any WIMPs that are within their radius of gravitational dominance. This implies that at $t_{\rm kd}$, the WIMPs have a uniform density distribution around the PBH.  The freeze-in picture is different since the FIMPs are always decoupled. If one assumes that the FIMPs thermalise within a hidden sector such that a well-defined temperature can be established, then there are cases in which the capture of FIMPs by PBH proceeds very differently.

The hidden sector can be thought of as initially cold and heated by energy transfer from the visible sector. Hence, a reasonable approximation for the ratio of hidden sector temperature $T'$ to visible sector temperature $T$ is given by \cite{Cheung:2010gj}
\beq
\frac{T'_{{\rm FI}}}{T_{{\rm FI}}} \sim \Big(M_{\rm Pl}  \langle \sigma v \rangle_{{\rm FI}} T_{{\rm FI}} \Big)^{1/4}~,
\eeq
where $\langle \sigma v \rangle$ is the production cross-section (in the IR or UV case), and evaluated at the time scale of freeze-in $t_{\rm FI}$. While WIMPs will be captured by PBH for  $m_{\rm DM} \gtrsim T$, for FIMPs one must instead compare to the hidden sector temperature, thus the condition is  $m_{\rm DM} \gtrsim T'$.

For UV freeze-in, FIMPs are produced immediately at the point of inflationary reheating and form a hidden sector bath at temperature $T'$. It follows that the history of halo formation is reminiscent of WIMPs, with the FIMPs free streaming until $m_{\rm DM} \gtrsim T'$, at which point they are captured by PBH. The only distinction is that one compares to the hidden sector temperature $T'$, rather than $T$.

The case of IR freeze-in is more complicated, as FIMPs are gradually populated over an extended period. The analysis of FIMP halo formation is simplified since we expect that, i).~the vast majority of FIMPs are produced at a characteristic time scale $t_{\rm FI}$; ii).~the infall time for particle dark matter is longer than the characteristic time scale for freeze-in: $t_{\rm FI}<t_{\rm collapse}$. 
In our analysis of the PBH halo, we will assume that at $t_{\rm FI}$ the FIMP density is uniform around the PBH. This assumption will be perturbed; (i) and (ii) above both fail to hold. In this case, FIMPs produced at early time will form a significant non-uniform density distribution, and at the later stages of freeze-in, one will have a uniform distribution of FIMPs {\em and} a significant spike profile. While this is not inconsistent, we find that (i) and (ii) typically do hold (see Appendix \ref{ApB}), and assuming an initially uniform FIMP distribution is reasonable.

\subsection{Density distribution of the dark matter halo at the time of collapse}

\label{s2.2}

 Let the time at which a PBH starts to capture dark matter particles (i.e. the time at which halo formation starts) around it be denoted as $t_{i}$. If $\Omega_{\rm DM}$ is set by the freeze-out mechanism, then we have $t_i=t_{{\rm kd}}$, the point in time when the dark matter particles get kinetically decoupled from the background plasma \cite{Boudaud:2021irr}. However, if $\Omega_{\rm DM}$ is fixed via the freeze-in mechanism, they are always assumed to be decoupled from the background plasma. If dark-matter particles produced during freeze-in are non-relativistic (as can be the case in IR freeze-in), the PBH gravitational field starts capturing them right from $t_i=t_{{\rm FI}}$. 
 
 Conversely, if the dark matter particles are relativistic at the time of production, they will free-stream out of the PBH gravitational well and evade capture. However, after freeze-in, the dark matter momenta redshift  $p \propto a^{-1} \propto T$ and becomes non-relativistic at some later time $t_{{\rm NR}}$ when the average dark matter momentum (or, equivalently, the dark matter temperature) becomes comparable $m_{\rm DM}$. In this case, we assume that the PBH starts capturing the dark matter particles at $t_i \sim t_{{\rm NR}}$. The temperature of the visible sector thermal bath at this point in time is then $T_{{\rm NR}}= m_{\rm DM} (T_{\rm FI}/T^{'}_{{\rm FI}})$. While analyzing specific models of dark matter freeze-in, we need to be careful about the difference between relativistic and non-relativistic freeze-in.

\newpage

For the PBH halo to begin to form, three conditions must be met; the majority of dark matter should have been produced, the dark matter should be non-relativistic, and the PBH should have formed. Generically, these conditions will be satisfied at different times (and not necessarily in the order above).
Hence, effective FIMP capture by the PBH will begin at 
\beq
t_i=\max[t_{\rm NR},~t_{\rm FI},~t_{\rm form}].
\label{ti}
\eeq
For PBH formed by the traditional process of collapsing horizon scale over-densities the formation time $t_{\rm form}$ (or the corresponding temperature $T_{\rm form}$) is given by \cite{Zeldovich:1967lct, Hawking:1971ei,Carr:1974nx,Carr:1975}\footnote{Alternatively, there is also the possibility of sub-horizon PBH formation, see e.g.~\cite{Ai:2024cka,Jung:2021mku,Belotsky:2014kca}, for which the formation time can be much later (and is not connected to $T_{\rm RH}$). We do not consider these cases here.}
\beq
t_{\rm form}\sim 10^{-5}~{\rm s} \left(\frac{M_\odot}{M_\bullet}\right)~, \quad \quad  T_{\rm form}\sim 1~{\rm GeV} \sqrt{\frac{M_\odot}{M_\bullet}}
\label{ffform}
\eeq
Observe that the time of formation is set by the mass of the PBH.

IR and UV freeze-in are distinct, so to maintain clarity, we discuss these separately.
\begin{itemize}
\item In the case of UV freeze-in one anticipates $t_{\rm FI}\sim t_{\rm RH}$ and $t_{\rm NR}> t_{\rm FI}$. Assuming PBH form due to horizon re-entry during radiation domination, generically $t_{\rm form}> t_{\rm RH}$, and we should take  $t_i=\max[t_{\rm NR},~t_{\rm form}]$. We note that these time scales depend on different physics, so either of $t_{\rm NR}$ or $t_{\rm form}$ could set $t_i$.

\item For IR freeze-in any of the three time scales ($t_{\rm NR},~t_{\rm FI},$ or $t_{\rm form}$) could set $t_i$. We can ask for what parameters  is $t_{\rm FI}>t_{\rm form}$, or equivalently $T_{\rm FI}<T_{\rm form}$, which give
\beq
\label{234}
\max(m_{\rm DM},m_M) \lesssim \sqrt{\frac{4\sqrt{90}}{\sqrt{g_*}} \frac{M_{\rm Pl}^3}{M_\bullet}}
\eeq 
where $m_M$ is the mediator mass.
When this is satisfied the time scale  $t_{\rm FI}$ is potentially relevant with $t_i=\max[t_{\rm NR},~t_{\rm FI}]$, then it remains to determine whether the dark matter is produced non-relativistically or not. Non-relativistic production occurs if the dark matter mass sets the freeze-in temperature, which is the case for $m_{\rm DM}>m_M$.
One must typically switch between $t_i=t_{\rm NR}$ and $t_i=t_{\rm FI}$ to analyse the full parameter space.
\end{itemize}

\noindent 
We note that it is possible that PBH may form prior to reheating (for instance, in an early matter-dominated phase) then $t_{\rm form}\lesssim t_{\rm RH}$. To allow for a larger possible range of possible PBH masses, we will also consider this scenario. However, we will clearly indicate when the parameter region is such that $t_{\rm form}< t_{\rm RH}$.

There is no appreciable capture prior to $t_i$, given in eq.~(\ref{ti}). One can define the radius of gravitational dominance of the PBH, the ``turn-around'' radius, given by $r_{{\rm ta}}(t)\simeq(2 G M_{\bullet} t^2)^{1/3}$, in terms of the mass of the PBH $M_{\bullet}$. At $t = t_i$, all of the dark matter particles that are within $r<r_i:=r_{\rm ta}(t_i)$ will decouple from the Hubble flow and start falling towards the PBH. 
Moreover, $r_{\rm ta}(t)$ increases with time and therefore the shell of dark matter particles between $r_i$ and $r_{\rm ta} (t)$ will also decouple from the Hubble flow and collapse towards the PBH. As the dark matter is nonrelativistic at $t>t_i$, it follows that $\rho_{\rm DM} \propto T^{3}$. Also, $r_{\rm ta} \propto T^{-4/3}$ during radiation domination, which implies the density of dark matter particles at radius $r$ when they start collapsing towards the PBH \cite{Boudaud:2021irr}
\begin{equation}\label{2.2}
    \rho_{\rm collapse} (r) = \begin{cases}
        \rho_i & r \leq r_i \\
        \rho_i (r/r_i)^{-9/4} & r_i<r \leq r_{\rm eq}\\
      0 &   r_{\rm eq}<r
    \end{cases}~,
\end{equation}
where $\rho_i=\rho(t_i)$ and we have introduced $r_{\rm eq}=r_{\rm ta}(t_{\rm eq})$, the turnaround radius at matter-radiation equality, which demarks the boundary of the PBH halo.  The profile of eq.~(\ref{2.2}) gives the starting density distribution, prior to subsequent evolution due to the gravitational pull of the PBH on the dark matter particles.

\subsection{FIMP halos around PBHs}
\label{sec2.2}

For $t>t_i$, the dark matter is non-relativistic and dark matter particles start collapsing towards a PBH.
 Whether a state remains captured or escapes the gravitational potential of the PBH depends on the velocity distribution at the time of the collapse. We approximate the velocity distribution of dark matter as a Maxwell-Boltzmann distribution with a velocity dispersion $\sigma_i \sim \sqrt{T'_i/m_{\rm DM}}$ where  $T'_i$ is the temperature of the dark matter particles at $t=t_i$. If dark matter is non-relativistic at freeze-in, then $T'_i= T'_{\rm FI} < m_{\rm DM}$, so $\sigma_i < 1$. Conversely, if the dark matter particles are relativistic at freeze-in, then $T'_i \sim m_{\rm DM}$ so $\sigma_i \sim 1$.

The velocity dispersion then redshifts as $\sigma \propto T$, and thus, the velocity dispersion of the dark matter particles at a radius $r$ from the PBH at the time of their collapse is
\begin{equation}\label{2.3}
    \sigma_{\rm collapse}(r) =  \begin{cases}
        \sigma_{i} & r \leq r_{i} \\
        \sigma_{i} (r/r_{i})^{-3/4} & r > r_{i}
    \end{cases}~.
\end{equation}
While the forms of eq.~(\ref{2.2}) \& (\ref{2.3}) match those derived in \cite{Boudaud:2021irr}, the quantity $t_i$, and thus the other quantities with subscripts, are very different between WIMPs and FIMPs.

With the initial density and velocity dispersion established in eq.~(\ref{2.2}) \& (\ref{2.3}), one can calculate the density profile of the PBH halo. As noted above, this leads to one of three possible broken power laws. The form of the broken power-law depends on the PBH mass and the particle dark matter \cite{Boudaud:2021irr}. The forms of the three cases are stated explicitly in Appendix~\ref{ApA}  (see \cite{Boudaud:2021irr} or \cite{Chanda:2022hls}).  As an instructive example, let us highlight that in the heavy PBH limit, in which case the halo evolves from eq.~(\ref{2.2}) to the following form
\begin{equation}\label{cases}
    \rho(r) \propto \begin{cases}
      r^{-3/2} &~~~~~ 0<r<r_1 \\
      r^{-9/4}  &~~~~~ r_1<r<r_{\rm eq} \\
        0 &~~~~~ r_{\rm eq}<r
    \end{cases}~.
    \end{equation}
At the time of formation,\footnote{Encounters with large astrophysical bodies can lead to significant stripping, thus truncating the particle dark matter halo \cite{Chanda:2022hls}. This is only relevant to PBH at late-time PBH (e.g.~galactic gamma-ray bounds), even then it is largely unimportant since essentially all gamma rays will be sourced by annihilations in the core. Thus, we will neglect this effect here.} the PBH halo terminates at $r_{\rm eq}$. Full forms for the above profiles, and the intermediate and light PBH cases, are given in Appendix \ref{ApA}.

The profile in eq.~(\ref{cases}) (as derived in \cite{Boudaud:2021irr} and stated in  Appendix \ref{ApA}) assumes that the particle dark matter has negligible interactions. Notably, particle dark matter annihilations can significantly impact the final density profile. The high density of dark matter particles in the halo's central region implies a high annihilation rate, resulting in a flattening of the central density spike  \cite{Chanda:2022hls,Adamek:2019gns,Gines:2022qzy,Kadota:2021jhg,Eroshenko:2016yve}.

To characterize the resultant flattened density of the central region, note that the annihilations will continue to decrease the central spike until the mean free time of a dark matter particle annihilating with another is equal to the age of the halo, i.e., $n \langle \sigma v \rangle \sim (t_{{\rm halo}})^{-1}$. Let the PBH be at a redshift $z$. If $t_0$ denotes the age of the universe and $H(z)$, $H_0$ represent the values of the Hubble parameter at redshift $z$ and at present, respectively, we have $t_{{\rm halo}}=t_0\times H_0/H(z)=t_0/h(z)$, and hence 
 \cite{Eroshenko:2016yve}: 
\begin{equation}
\rho_{{\rm core}} (z) \sim \frac{m_{\rm DM}}{\langle \sigma v \rangle t_{{\rm halo}}} = \frac{m_{\rm DM} h(z)}{\langle \sigma v \rangle t_0}.
\end{equation}
For s-wave annihilating dark matter,\footnote{For an examination of annihilation processes around PBH beyond s-wave, see \cite{Chanda:2022hls,Kadota:2021jhg}.} the central region is thus a constant density core (plateau), whereas, for dark matter models with velocity-dependent cross-sections, $\rho_{{\rm core}}$ it is no more constant but inherits a radial dependence via $v=\sqrt{\frac{2 G M_{\bullet}}{r}}$.

\begin{figure}[t] 
\centerline{    \includegraphics[scale=0.62]{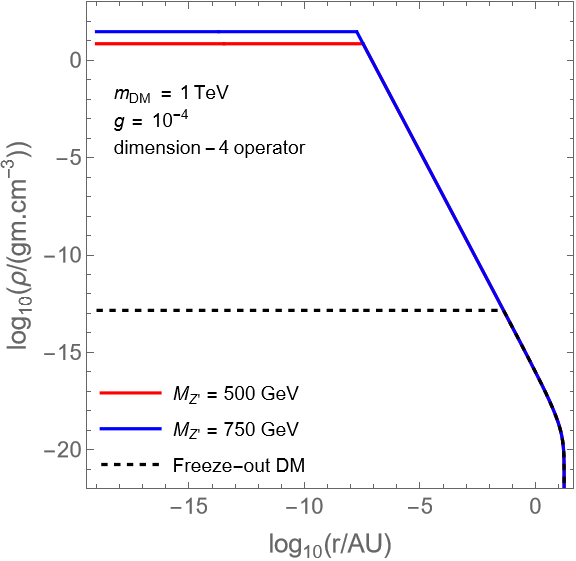} \hspace{6mm}
    \includegraphics[scale=0.62]{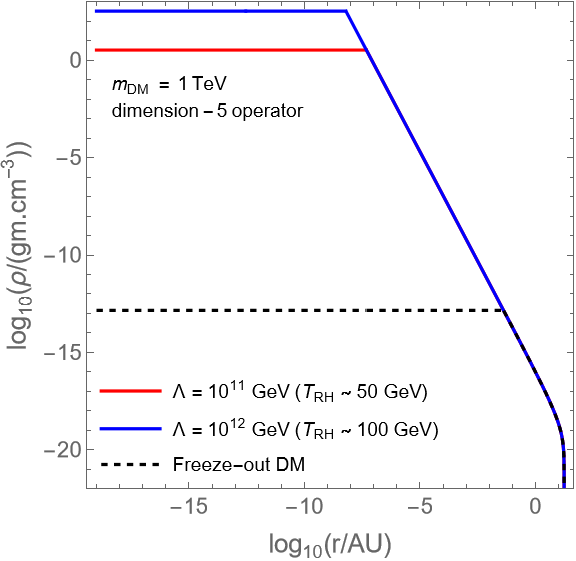} }
   \caption{Halo profiles for 1 TeV dark matter around a PBH of mass $M_{\bullet} = 10^{-6} M_{\odot}$. The left panel represents the halo profile for dark matter for the dimension 4 freeze-in model for two values of the mediator mass $M_{Z'} = 500$ GeV and $M_{Z'} = 750$ GeV, and a DM-$Z'$ coupling value of $g = 10^{-4}$. The right panel represents the same for a dimension-5 UV freeze-in model via a non-renormalizable operator. We use two values of $\Lambda$ (the energy scale of new physics for UV freeze in) with the corresponding $T_{\rm RH}$ values shown as well. In both panels, halo profiles for freeze-out dark matter (see e.g.~\cite{Chanda:2022hls}) are shown as dashed lines for comparison. Observe that the central plateau region is much smaller than freeze-out dark matter for freeze-in dark matter in both panels.}
\label{fig:halo}
\end{figure}

As $\rho(r)$ decreases rapidly with $r$, we will eventually find that beyond a distance $r = r_{\rm core}$ that $\rho(r) < \rho_{\rm core}$. Hence, the rate of dark matter annihilations is too low to have any significant effect in further reducing the halo density at that distance. Thus, the final density profile for the dark matter halo around a PBH will be given by the reduced central density profile $\rho_{{\rm core}}$ in the region $0<r<r_{{\rm core}}$, which then smoothly transitions into one of the analytic forms of $\rho(r)$ given in Appendix \ref{ApA} at $r = r_{{\rm core}}$.

We can determine the size of the central region depleted by annihilations $r_{\rm core}$  by matching the core density $\rho_{{\rm core}}$ to the innermost profile derived in Section \ref{s2.2} (and Appendix \ref{ApA}). In Figure~\ref{fig:halo} (right) we compare the WIMP and FIMP halo profiles for the case that $m_{\rm DM}=1$ TeV reproduces 99\% of the correct $\Omega_{\rm DM}$ and assuming a PBH of mass $10^{-6} M_{\odot}$.

Since freeze-in and freeze-out mechanisms imply very different strength interactions for the dark matter particles, it is unsurprising that their annihilation cross-sections are significantly different. As an instructive example, let us consider IR freeze-in with characteristic coupling $\lambda\sim10^{-10}$  \cite{Hall:2009bx} and take $m_{\rm DM}\sim 1$ TeV, thus s-wave annihilation cross-section 
\beq
\langle \sigma v \rangle_{\rm FIMP} \sim \frac{\lambda^2}{m_{\rm DM}^2}\sim 10^{-26}~{\rm GeV}^{-2} \left(\frac{\lambda}{10^{-10}}\right)^2\left(\frac{1~{\rm TeV}}{m_{\rm DM}}\right)^2.
\eeq
 If we compare this to a WIMP with  $
\langle \sigma v \rangle_{\rm WIMP} \approx 3 \times 10^{-26} \, \text{cm}^3/\text{s} \simeq  10^{-9} \, \text{GeV}^{-2}$, being the standard thermal cross-section, we see that the highest density in the PBH halo is many orders of magnitude higher for FIMPs (compared to WIMPs of comparable mass)
\beq
\left. \frac{\rho_{\rm FIMP}}{\rho_{\rm WIMP}} \right|_{\rm core}\sim  \frac{\langle \sigma v \rangle_{\rm WIMP}}{\langle \sigma v \rangle_{\rm FIMP}}\sim 
10^{17} \left(\frac{10^{-10}}{\lambda}\right)^2\left(\frac{m_{\rm DM}}{1~{\rm TeV}}\right)^2.
\eeq
This, in turn, implies that the plateau region for FIMP halos will be much smaller than for WIMPs, or even non-existent (i.e.~smaller than the PBH Schwarzschild radius $r_s=2GM_{\bullet}$).  As can be seen in the panels of Figure~\ref{fig:halo}. This plateau follows from the assumption of a velocity-independent annihilation cross-section. In the case of velocity-dependent annihilations, the late-time density profile will be quite different \cite{Kadota:2021jhg,Gines:2022qzy,Chanda:2022hls}, although this work will restrict our study to the velocity-independent case.

\section{Indirect Detection Bounds}
\label{S3}

We next derive the annihilation rate of FIMPs around the PBH and, subsequently, the exclusion limits. These depend on the density profiles of the previous section. Specifically, we mainly focus on the Fermi-LAT observations of the total extragalactic $\gamma$-ray flux \cite{Fermi-LAT:2014ryh} and our limits are derived by requiring that  FIMP annihilations in the total population of PBH halos do not exceed the current observed flux.

We highlight that complementary constraints already exist on PBH (without reference to dark matter particles), most prominently from searches for microlensing, and we shall incorporate these into our exclusion plots.  An important restriction comes from PBH evaporation due to Hawking radiation \cite{Hawking:1974rv}, a population lighter than $5\times10^{14}$ g should have entirely evaporated by today. Moreover, PBH actively evaporating presently leads to strong astrophysical constraints (see e.g.~\cite{Carr:2020gox,Khlopov:2008qy}). These PBH limits already fairly strongly constrain the parameter space without reference to dark matter. Notably, PBH can only be 100\% of the dark matter in a small ``asteroid mass'' window ($10^{-10} M_\odot\lesssim M_{\bullet}\lesssim 10^{-7} M_\odot$), for $M_{\bullet}\sim1-10 M_\odot$ PBH can be 1-10\%, but outside of these two ranges the PBH fraction must be sub-percent level. These limits assume non-rotating, non-charged PBH with monochromatic mass spectra, and deviations from these assumptions do alter these PBH constraints.

\subsection{Gamma-ray flux from FIMP annihilations}

To calculate the annihilation rate $\Gamma_{\rm PBH}$ in a dark matter halo, one integrates over the density profile $\rho(r)$ as follows \cite{Cirelli:2010xx}
\begin{equation}\label{eq:AnnihilationRate0}
\Gamma_{{\rm PBH}}(z) = 4\pi \int dr~r^{2}\left(\frac{\rho(r,z)}{m_{\rm DM}}\right)^{2}\langle\sigma v\rangle~.
\end{equation}
Such dark matter particle annihilations in the PBH halo can result in the production of ${\gamma}$-rays. The density profile exhibits a $z$ dependence due to the fact that young low-$z$ PBH halos will not have depleted their inner density cores through dark matter particle annihilation, in contrast to late-time high-$z$ PBH halos. Analytic forms for the annihilation rate, including the full $z$-dependence, are given in  \cite{Chanda:2022hls}.

The total power density thus radiated  at a specific energy $E'_{\gamma}$, per unit energy interval, from the PBH halos situated at redshift $z$, is given by the $j$-factor
\begin{equation}
    j(E'_{\gamma}, z) \simeq  \Gamma_{\rm PBH}(z)  n_{\rm PBH}(z) \left[E \frac{{\rm d}N}{{\rm d}E} \right]_{E=E'_{\gamma}}~.
\end{equation}
The number density of PBH  is $n_{\rm PBH}|_{\rm today} = f_{\rm PBH} 
\rho_{{\rm DM},0} = \Omega_{{\rm DM},0} \rho_c  /M_{\bullet}$ and redshifts such that $ n_{\rm PBH}(z) = (1+z)^3 \times n_{\rm PBH}|_{\rm today} $. The differential photon spectrum   $E \frac{dN}{dE}|_{E=E'_{\gamma}}$ is comprised of the photons from annihilations within PBH halos at all (observable) redshifts. While the details of the photon spectrum are model-dependent, one can make reasonable estimates by assuming a specific spectrum or by taking a specific particle physics model. Note that the produced photons will reach Earth with a red-shifted energy according to $E_{\gamma}=E'_{\gamma}/(1+z)$. 
The observed differential flux due to these $\gamma$-ray photons is given by
\begin{equation}\label{flux}
    \frac{{\rm d}^2\Phi(E_{\gamma})}{{\rm d}E {\rm d}\Omega} = \frac{1}{4 \pi E_{\gamma}} \int \frac{dz}{H(z) (1+z)^4} j(E'_{\gamma}, z) e^{-\tau(E_{\gamma},z)}~,
\end{equation}
where the optical depth parameter for absorption of extra-galactic $\gamma$-rays is denoted by $\tau$. We use the photon spectra and optical depth function provided in \cite{Cirelli:2010xx,Elor:2015bho} for our calculations.

For a given scenario, the predicted flux can then be calculated and compared with the Fermi-LAT observations of the total extragalactic $\gamma$-ray flux \cite{Fermi-LAT:2014ryh} to determine if there is any constraint on the value of $f_{\rm PBH}$ for a given $M_{\bullet}$ (assuming a monochromatic mass spectrum for the PBHs). Specifically, we calculate our constraint to be the value of $f_{\rm PBH}$ at which the flux from FIMP annihilations saturates the total extragalactic $\gamma$-ray background observed by Fermi-LAT at any energy. In what follows, we will study the case in which a population of PBH is accompanied by an abundance of dark matter particles, considering first a standard example of a renormalisable (IR) freeze-in model and subsequently generic dimension-5 and dimension-6 UV freeze-in operators.

Specifically, in what follows, we will consider the case of dark matter annihilations mediated via a $Z'$ to fermions (which will specify the annihilation cross section), and we then further specialise to assume the $Z'$ only decays to $b\bar{b}$-pairs. Once the $b$-quarks are produced, these will hadronize and this process leads to photons. This hadronisation process depends on the energy of the $b$-quark, which depends on the mass of the dark matter. The hadronisation process of $b$-quarks is well understood and can be described by a differential photon spectrum which is tabulated in e.g.~\cite{Cirelli:2010xx,Elor:2015bho}. We obtain limits specifically for this scenario, although we expect mild variations in mediator types and/or decay channels to be qualitatively similar.\footnote{For a more complicated scenario in which the $Z'$ decays to states other than bottom quarks, each decay product will have a distinct photon spectrum. A previous paper \cite{Chanda:2022hls} for WIMP annihilations around PBHs explored $Z'$  decays to other final states, specifically $\tau\bar{\tau}$ and $W^+W^-$, and found that while there are differences, these lead to qualitatively similar results (when viewed on a log plot).}

\subsection{IR freeze-in} \label{4.1}

For our example of renormalisable freeze-in, we suppose that the Standard Model is supplemented by two new states: dark matter fermions $X$ with mass $m_{\rm DM}$ and a $Z'$ mediator with mass $M_{Z'}$ connecting the dark matter states with Standard Model fermions $f$. The terms in the Lagrangian governing the $Z'$ interactions with the dark matter sector and the visible sector are given by the renormalizable operators $g Z' \bar{X} X$ and $\lambda Z' \bar{f} f$, respectively. Depending on the hierarchies of the masses and couplings, we can have two different scenarios:\footnote{We neglect the case in which both these processes are comparable.} 
\begin{enumerate}
\item[\bf A.] Freeze-in is primarily via $f \bar{f} \rightarrow X \bar{X}$ for $M_{Z'} < 2 m_{\rm DM}$ or if $M_{Z'} > 2 m_{\rm DM}$ and $g \ll \lambda$.
\item[\bf B.] If $M_{Z'} > 2 m_{\rm DM}$ and $\lambda \ll g$, we first have freeze in of $Z'$ via $f \bar{f} \rightarrow Z'$, followed by the decay of $Z'$ to $X \bar{X}$ pairs, as in \cite{Scholtz:2019csj}.
\end{enumerate}

\vspace{3mm}\noindent {\bf Case A} is dominated by the 2$\rightarrow2$ process, to find the relic density, we first write the Boltzmann equation governing the evolution of number density $n_{X}$
\begin{equation} \label{eq:4.4}
    \frac{dn_{X}}{dt} + 3 H n_{X} = \int d\Pi_f d\Pi_{\bar{f}} d\Pi_{\rm DM} d\Pi_{\bar{X}} (2 \pi)^4 \delta^4(p_{\rm DM} + p_{\bar{X}} - p_f - p_{\bar{f}}) |\mathcal{M}|^2_{f \bar{f} \rightarrow X \bar{X}} f^{\rm eq}_{f} f^{\rm eq}_{\bar{f}}.
\end{equation}
Neglecting the Standard Model fermion masses, the matrix element for the 2$\rightarrow$2 process is
\begin{equation}
\label{MM}
 |\mathcal{M}|^2_{f \bar{f} \rightarrow X \bar{X}} \simeq  g^2 \lambda^2 \frac{s(s-4m^2_{\rm DM})}{(s - M_{Z'}^2)^2+M_{Z'}^2 \Gamma_{Z'}^2}~,
\end{equation}
where $\Gamma_{Z'}$ is the decay width of $Z'$.  There are two subcases depending on the mass hierarchy.

\vspace{3mm}\noindent {\bf Case A.i} For $M_{Z'} \ll 2 m_{\rm DM} < \sqrt{s}$, at high temperatures we can approximate $|\mathcal{M}|^2 \simeq g^2 \lambda^2$. 
With this form of the matrix element, we evaluate the integral of eq.~(\ref{eq:4.4}) to arrive at the yield $Y_X$. Specifically, for $M_{Z'} < m_{\rm DM}$, we find
\begin{equation} \label{eq:3.6}
    Y_{X} = \frac{n_{X}}{s} \approx \frac{135 g^2 \lambda^2 M_{\rm Pl} }{2^{13} \pi^5 g_*^S \sqrt{g_*^{\rho}} (1.66) m_{\rm DM}}~,
\end{equation}
and it follows that the relic abundance at present is given by
\begin{equation} \label{eq:3.7}
    \Omega_{\rm FIMP}h^2 \simeq  \left(\frac{2 \times 10^{23}}{g_*^S \sqrt{g_*^\rho}}\right) g^2 \lambda^2
    \simeq0.1  \left(\frac{g\lambda}{3\times10^{-11}}\right)^2~.
\end{equation}

\vspace{3mm}\noindent {\bf Case A.ii} In the converse case with $M_{Z'} > 2 m_{\rm DM}$, a $Z'$ resonance occurs for $s\approx M_{Z'}^2$ and using the narrow-width approximation, we can express the squared matrix element as 
\beq
 |\mathcal{M}|^2_{f \bar{f} \rightarrow X \bar{X}} \simeq  g^2 \lambda^2 \frac{s(s-4m^2_{\rm DM})}{M_{Z'} \Gamma_{Z'}} \pi \delta(s-M_{Z'}^2)~,
\eeq 
where we have neglected the masses of Standard Model fermions and the dark matter particle, and take $\Gamma_{Z'} \simeq \frac{\lambda^2 + g^2}{12 \pi} M_{Z'}$. The equilibrium densities of the Standard Model bath particles $f$ are taken to be $f^{\rm eq} \approx e^{-E/T}$.
With this form of the matrix element, for $M_{Z'} > m_{\rm DM}$, it follows from eq.~(\ref{eq:4.4}) that
\begin{equation} \label{eq:3.9}
    Y_{X} = \frac{n_{X}}{s} \approx \frac{405 g^2 \lambda^2 M_{\rm Pl}}{512 \pi^4 g_*^S \sqrt{g_*^{\rho}} (1.66) M_{Z'} (\lambda^2 + g^2)}.
\end{equation}
Taking $\lambda \gg g$ (thus $Z'$ is effectively in the FIMP sector),  the relic abundance is given by 
\begin{equation} \label{eq:3.10}
     \Omega_{\rm FIMP}h^2 \simeq 0.1  \times \left( \frac{g}{10^{-11}} \right)^2\left( \frac{30~{\rm TeV}}{M_{Z'}} \right)\left( \frac{m_{\rm DM}}{1~{\rm TeV}} \right)~.
\end{equation}
Since  $\Omega_{\rm DM}h^2\approx 0.1$, provided that\footnote{\label{foot} We neglect the case that $f_{\rm PBH}\approx1$ and thus the particle dark matter is subdominant to the PBH population. This is justified, given the strength of the microlensing limits over most of the PBH mass range.} $f_{\rm PBH}\not\approx0.1$, the couplings can thus be determined by matching the observed dark matter relic abundance.

\vspace{3mm}\noindent {\bf Case B} is distinct since the FIMPs are produced via the two-step process $f\bar{f}\rightarrow Z'$ controlled by $\lambda$, followed by the (potentially long-lived) decay $Z'\rightarrow X\bar{X}$  controlled by $g$. For $\lambda\ll g$, the FIMP abundance is set by the freeze-in abundance of $Z'$ via $Y_{X} \simeq Y_{Z'}/2$ since the branching fraction back to the visible sector is small. As such, the coupling $g$ does not play a major role in this freeze-in scenario, provided  $\lambda\ll g$ and that $g$ is sufficiently small to avoid sector thermalization (we also assume $g\lesssim1$ to avoid strong FIMP self-interactions that would impact the halo profile).

Considering the process $f\bar{f}\rightarrow Z'$, the Boltzmann equation for evolution of $n_{Z'}$ is  \cite{Hall:2009bx}
\begin{equation}
    \frac{dn_{Z'}}{dt} + 3 H n_{Z'} = \int d\Pi_f d\Pi_{\bar{f}} d\Pi_{Z'} (2 \pi)^4 \delta^4(p_{Z'} - p_f - p_{\bar{f}}) |\mathcal{M}|^2_{f \bar{f} \rightarrow Z'} f^{\rm eq}_{f} f^{\rm eq}_{\bar{f}}~.
\end{equation}
and $|\mathcal{M}|^2_{f \bar{f} \rightarrow Z'} \sim \lambda^2$. Solving the differential equation, we will get the freeze-in abundance $Y_{Z'}$, which then decays to give the FIMP relic density. Assuming $m_f \ll m_{\rm DM} < M_{Z'}$ the present dark matter relic abundance is parametrically \cite{Scholtz:2019csj}
\begin{equation} \label{eq:3.12}
    \Omega_{\rm FIMP} h^2\simeq 0.1 \times \left(\frac{\lambda}{3 \times 10^{-12}} \right)^2 
\left(\frac{30 \text{ TeV}}{M_{Z'}} \right) \left(\frac{m_{\rm DM}}{1 \text{ TeV}}\right)~.
\end{equation}
So in this case $\lambda$, as a parameter, is fully determined by $m_{\rm DM}$ and $M_{Z'}$ and is independent of $g$. Also, the temperature of the universe at the time of freeze-in is $T_{\rm FI} \sim M_{Z'}$.

Having understood the relations between parameters needed for $\Omega_{\rm FIMP}\simeq\Omega_{\rm DM}\simeq0.1$, we can use these parameters to calculate the form of the FIMP annihilation cross-section.
While the FIMPs do not undergo annihilations at early time (during freeze-in), at late time, FIMP annihilations can potentially occur in regions of high density. Conventionally, such annihilations will also be generically absent due to the small annihilation cross-section. However, since PBHs can lead to extremely high FIMP densities, this revives the prospect of indirect detection signals.
For $M_{Z'} > m_{\rm DM}$, dark matter annihilations can only occur through the s-channel process $X \bar{X} \rightarrow f \bar{f}$. On the other hand, if $M_{Z'} < m_{\rm DM}$, in addition to -s-channel annihilations, dark matter can also annihilate through the t-channel $X \bar{X'} \rightarrow Z' Z'$, followed by $Z' \rightarrow f \bar{f}$. Assuming we are away from the $Z'$ resonance, the non-relativistic expressions for thermally averaged cross-sections are \cite{Alves:2015pea}
\begin{equation} \label{eq:3.13}
     \langle \sigma v \rangle_{X\bar{X}\rightarrow f \bar{f}} = \frac{\lambda^2 g^2 N_c}{2 \pi} \sqrt{1-\frac{m_f^2}{m_{\rm DM}^2}} \left\{\frac{4 m_{\rm DM}^2-m_f^2}{(4 m_{\rm DM}^2 - M_{Z'}^2)^2} + \frac{m_f^2}{M_{Z'}^4} \right\},
\end{equation}
and the t-channel process
\begin{equation} \label{eq:4.10}
    \langle \sigma v \rangle_{X\bar{X}\rightarrow Z' Z'} = \frac{g^4}{4 \pi M_{Z'}^2} \left(1-\frac{M_{Z'}^2}{m_{\rm DM}^2}\right)^{3/2} \left(1-\frac{M_{Z'}^2}{2 m_{\rm DM}^2}\right)^{-2} \left(2-\frac{M_{Z'}^2}{m_{\rm DM}^2}\right).
\end{equation}
We shall restrict our attention here\footnote{The converse case is interesting but distinct and thus we will return to it in a future publication.} to the case that  $m_f\ll m_{\rm DM}, M_{Z'}$. In the case that $m_{\rm DM} \gg M_{Z'}$ annihilations can proceed through both the s-channel process with cross-section
\beq 
\langle \sigma v \rangle_{X\bar{X}\rightarrow f \bar{f}}  &\simeq \frac{N_c g^2 \lambda^2}{2 \pi m_{\rm DM}^2} \quad,\quad
\eeq
as well as t-channel processes with
\beq
\langle \sigma v \rangle_{X\bar{X}\rightarrow Z' Z'}     &\simeq  \frac{g^4}{4 \pi M_{Z'}^2}~.
\eeq
In such cases, the dominant channel of annihilation has to be determined by looking at the ratio of the two annihilation cross-sections. If $\frac{ \langle \sigma v \rangle_{X\bar{X}\rightarrow Z' Z'} }{ \langle \sigma v \rangle_{X\bar{X}\rightarrow f \bar{f}} } \gg 1$, t-channel dominates over s-channel as the primary annihilation route, and vice versa. 
For example, when $m_{\rm{DM}} \gg M_{Z'}$, eq.~(\ref{eq:3.13}) and eq.~(\ref{eq:4.10}) imply that the t-channel is dominant over the s-channel provided
\begin{equation}
    \left( \frac{g}{\lambda} \right)^2 \gg \frac{2 N_c M_{Z'}^2}{m^2_{\rm{DM}}}~.
\end{equation}

\subsection{Limits on the PBH-FIMP scenario for IR freeze-in}

Using the density profiles from Section \ref{sec2.2} and the cross-sections in eq.~(\ref{eq:3.13}) and eq.~(\ref{eq:4.10}), we can calculate the annihilation rate and the flux via  eq.~(\ref{eq:AnnihilationRate0}) \& (\ref{flux}).  The extragalactic $\gamma$-ray flux has been measured by Fermi-LAT. Using eq.~(\ref{flux}), we can compute flux due to annihilations in the PBH halo as a function of PBH mass $M_{\bullet}$, FIMP mass $m_{\rm DM}$, the $Z'$ mass $m_{Z'}$, the couplings $g$ and $\lambda$, and the fractional abundance. We will fix the coupling $\lambda$ via the condition that freeze-in leads to the observed dark matter relic density, making the assumption that $\Omega_{\rm FIMP}\simeq\Omega_{\rm DM}\gg \Omega_{\rm PBH}$ (cf.~Footnote \ref{foot}).

\begin{figure}[t] 
    \centerline{
    \includegraphics[scale=0.67]{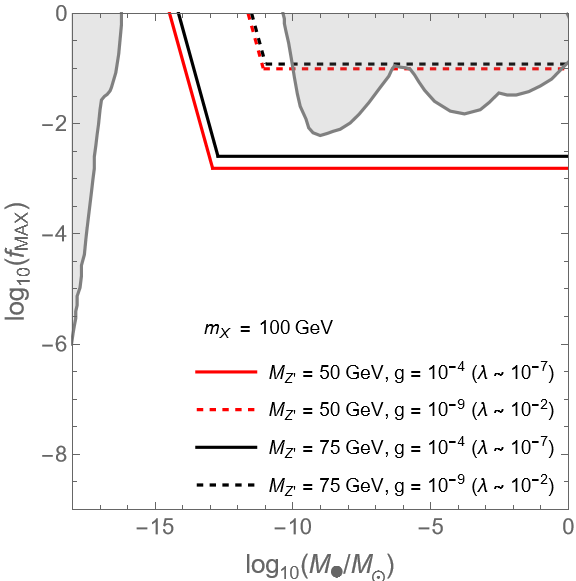} \hspace{1cm}
    \includegraphics[scale=0.67]{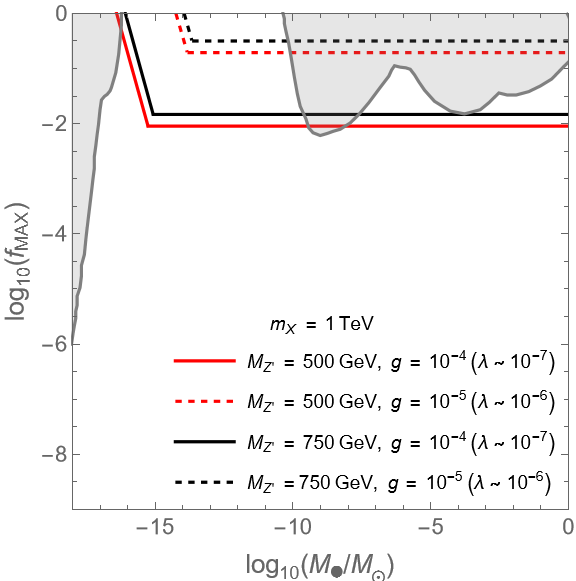} }
    \caption{{\bf Dimension 4.} We assume FIMP freeze-in via a  $Z'$ with $M_{Z'} < 2 m_{\rm DM}$, involving a renormalisable interaction.  The plots show the maximum PBH fractional abundance consistent with Fermi-LAT extragalactic $\gamma$-ray background, as a function of PBH mass $M_{\bullet}$. 
Since $M_{Z'} < 2 m_{\rm DM}$, in both panels the primary freeze-in channel is via $f \bar{f} \rightarrow X \bar{X}$. Both panels correspond to {\bf Case A.i} discussed in the text. The combination $g\lambda$ sets the relic abundance, cf.~eq.~(\ref{eq:3.7}).
The grey regions are the standard constraints on PBH coming from evaporations, gravitational waves, lensing, and CMB distortions (see e.g.~\cite{Carr:2020gox}).
{\bf Left Panel.} For a FIMP-$Z'$ coupling value of $g = 10^{-4}$ (solid lines), we have t-channel annihilation $X \bar{X} \rightarrow Z'Z' \rightarrow b \bar{b} b \bar{b}$ as the dominant annihilation channel for both values of $M_{Z'}$, while
for the lines corresponding to $g = 10^{-9}$ (dashed), the s-channel annihilation $X \bar{X} \rightarrow b \bar{b}$ is dominant. 
 {\bf Right Panel.} For $m_{\rm DM} = 1 {\rm ~ TeV}$, we only get constraints on $f_{{\rm PBH}}$ when the t-channel annihilation is dominant, i.e. when $M_{Z'} < m_{\rm DM}$ and $g \lesssim 10^{-5}$. In both panels, we check that the couplings are not so large that the FIMPs come into equilibrium with the visible sector. Specifically, for the values of the $Z'$ mass we find that avoiding equilibration restricts  $g \lesssim 10^{-3}$ (with $\lambda\sim10^{-11}/g$).}
\label{fig:dim4}
\end{figure}

 Notably, for a given PBH mass, dark matter mass, and annihilation cross-section, one can identify a maximum
fractional abundance $f_{\rm MAX}$ such that  $f_{\rm PBH} > f_{\rm MAX}$ would imply the flux due to PBH would exceed the observed total flux at some energy.  More details on our analysis are given in Appendix \ref{ApD} alongside a discussion of some differences with the analysis in \cite{Chanda:2022hls} and a comparison with earlier papers  \cite{Chanda:2022hls,Gines:2022qzy}. Thus $f_{\rm MAX}$ places a limit on the maximum size of the PBH fraction for a given PBH mass and FIMP model. We show the limit on $f_{\rm PBH}$ as a function of PBH mass in Figure~\ref{fig:dim4} with different assumptions regarding the IR freeze-in model. In all cases, we assume the $Z'$ primarily decays to $b\bar{b}$ pairs of the Standard Model states.
We note that dark matter models for which t-channel annihilation $X\bar{X}\rightarrow Z'Z'$ (i.e.~for $m_{Z'}<m_{\rm DM}$) is possible are strongly constrained, whereas models with only s-channel annihilation  $X\bar{X}\rightarrow b\bar{b}$ are very loosely unconstrained. We take care to ensure that the couplings are such that the FIMPs will thermalise with the visible sector.

\begin{figure}[t] 
    \centerline{
    \includegraphics[scale=0.67]{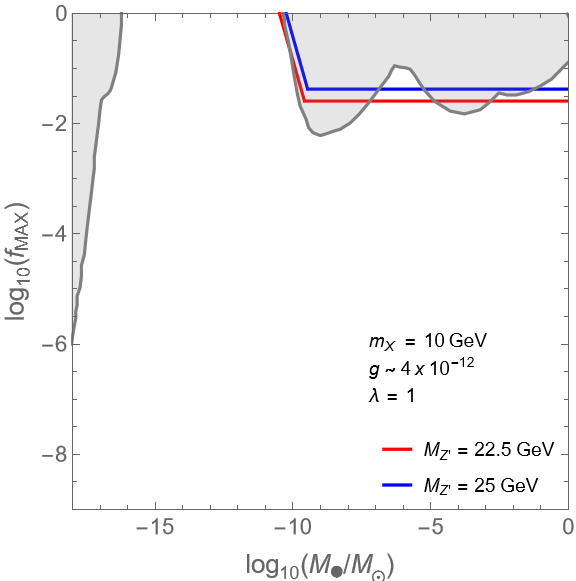}  \hspace{1cm}
    \includegraphics[scale=0.67]{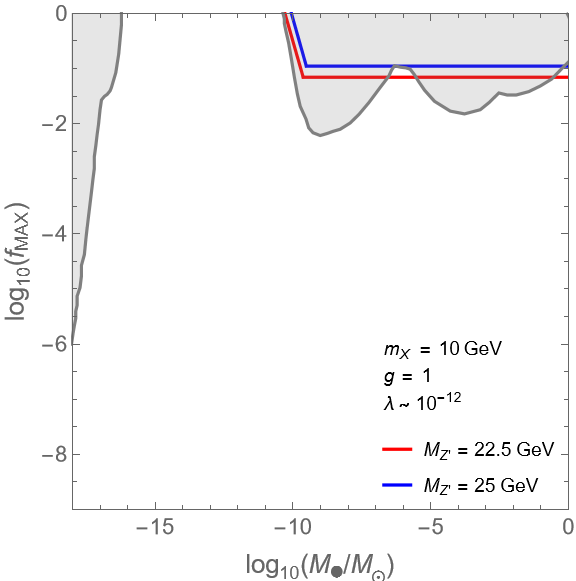} }
    \caption{{\bf Dimension 4.} As Figure \ref{fig:dim4} but with $M_{Z'} > 2 m_{\rm DM}$.
     {\bf Left [Case A.ii].} Freeze-in  occurs via $f \bar{f} \rightarrow X \bar{X}$, the FIMP relic density is determined by eq.~(\ref{eq:3.10}) which fixes $g$. {\bf Right [Case B].} Freeze-in of $X$ occurs via the two-step process $f \bar{f} \rightarrow Z'$ and $Z'\rightarrow X \bar{X}$, the FIMP relic density is set by eq.~(\ref{eq:3.12}) which fixes $\lambda$. In both cases, one coupling is fixed by the relic abundance, and we are free to choose the other. We confirm that the parameters taken do not lead to sector equilibration.}
\label{fig:dim4m10}
\end{figure}

The limits for Case A.i, described above, are shown in Figure~\ref{fig:dim4}. For cases A.ii and B, i.e.~when $M_{Z'} > 2m_{\rm DM}$, we only get limits when the dark matter mass is light. In Figure~\ref{fig:dim4m10} we compute the limits on $f_{\rm PBH}$ for a 10 GeV FIMP for Case A.ii (left panel) and Case B (right panel). We find that both cases A.ii and B are only loosely unconstrained for 10 GeV dark matter, with the constraints on annihilations in the PBH halo being largely less constraining than the model-independent PBH microlensing limits shown in grey.

We highlight at this point that if the couplings are too large, then the FIMPs will thermalise with the visible sector, spoiling the freeze-in mechanism (details can be found in Appendix \ref{ApC}). We are careful to ensure that for the parameter values given in Figures \ref{fig:dim4} \& \ref{fig:dim4m10} that the couplings are sufficiently small that the FIMPs do not equilibrate with the Standard Model thermal bath. Specifically, as derived in Appendix \ref{ApC},  we find that for Case A.i with $m_{\rm DM}\sim100$ GeV - 1 TeV and $m_{\rm DM}/m_{Z'}\sim2$ then we require  $g \lesssim 10^{-3}$. This assumes that  $g\lambda\sim10^{-11}$, which is the value of the combination of couplings required such that the FIMPs account for the observed dark matter relic density, as can be seen in eq.~(\ref{eq:3.7}). 
Further, for Case A.i and Case B, taking $M_{Z'}\sim25$ GeV, as assumed in Figure \ref{fig:dim4m10}, we find that sector equilibration is avoided for $g\ll5\times10^{-8}$ and $\lambda\ll3\times10^{-8}$ respectively (cf.~Appendix \ref{ApC}).

\subsection{UV freeze-In}

In models of UV freeze-in, the portal operator that connects the FIMPs to the visible sector are non-renormalisable operators.
As a result, UV freeze-in is fundamentally different from its renormalisable counterpart, as the freeze-in yield depends polynomially on the highest temperature of the standard model thermal bath $T_{\rm RH}$. 
Unlike IR freeze-in, freeze-in production via non-renormalizable higher-dimensional operators is UV-dominated at the highest temperature scales of the universe.
In the context of effective field theory, the higher-dimensional operator can be viewed as a low-energy approximation of some new physics at a higher energy scale $\Lambda$. The thermally averaged cross-section of the 2-to-2 scattering mediated by the higher-dimensional operator can be expressed as
\begin{equation}
    \langle \sigma v \rangle \simeq \frac{T^n}{\Lambda^{n+2}}~.
\end{equation}
For dimension~5 operators such as $\frac{1}{\Lambda} \phi^{\dagger} \phi \bar{f} f$ involving Standard Model fermions $f$ and scalar FIMP $\phi$ one takes $n=0$. The portal operator is dimension~6 in the case of fermion FIMP $X$ with the form $\frac{1}{\Lambda^2}  \bar{X}X  \bar{f} f$, corresponding to $n=2$.
Provided that $m_{\rm DM}\ll T_{\rm R}$ (and for consistency $T_{\rm RH}<\Lambda$) then UV freeze-in leads to a yield which scales as (cf.~\cite{Elahi:2014fsa})
\beq\label{UVY}
Y_{\rm FIMP}\propto\frac{M_{\rm Pl} T_{\rm RH}^{2n+1}}{\Lambda^{2n+2}}~.
\eeq
For ``vanilla'' UV freeze-in which $m_{\rm DM}<T_{\rm RH}$ the parameters are simply $m_{\rm DM}$,  $\Lambda$, and $T_{\rm RH}$.
Further, the condition $\Omega_{\rm FIMP}\simeq\Omega_{\rm DM}$ provides a relationship between the FIMP parameters, and we can also calculate the annihilation cross-section for FIMPs via the same portal operator.  By dimensional analysis, the annihilation cross-section for scalar FIMPs $\phi$ into Standard Model fermions (involving the dimension five operator) is given by
\begin{equation} \label{eq:3.17}
    \langle \sigma v \rangle_{\phi \phi^{\dagger} \rightarrow f \bar{f}} \approx \frac{1}{\Lambda^2}~.
\end{equation}

The  annihilation cross-section for fermion FIMPs $X$ into Standard Model fermions (involving the dimension five operator) is given by
\begin{equation} \label{eq:3.18}
\langle \sigma v \rangle_{X \bar{X} \rightarrow f \bar{f}} \approx \frac{m_{\rm DM}^2}{\Lambda^4}~.
\end{equation}
With the ``vanilla'' UV freeze-in with $m_{\rm DM}<T_{\rm RH}$, we find there are no limits from indirect detection sources by annihilations in the  FIMP halo of the PBHs. However, there is a distinct class of models in which $m_{\rm DM}>T_{\rm RH}$ for which limits do arise as we discuss next.

\newpage

\vspace{-2cm}

\clearpage

\subsection{Boltzmann Suppressed Freeze-in}
\label{superuv}

In a pioneering paper 25 years ago, Giudice, Kolb, \& Riotto  \cite{Giudice:2000ex} explored many general scenarios of (what is now called) freeze-in, including freeze-in with $m_{\rm DM}>T_{\rm RH}$. In the IR case, this has been referred to as ``Freeze-in at stronger coupling'' \cite{Cosme:2023xpa,Arcadi:2024obp,Bernal:2024ndy,Belanger:2024yoj,Boddy:2024vgt}. The UV case has been largely ignored outside of \cite{Giudice:2000ex}, however, we highlight the recent paper \cite{Bernal:2025xx}. We shall refer to both IR and UV cases with $m_{\rm DM}>T_{\rm RH}$ as ``Boltzmann Suppressed Freeze-in''.

Of particular relevance, the authors of \cite{Giudice:2000ex} highlighted that the highest temperature reached by the visible sector thermal bath $T_{\rm max}$ can be higher than the conventionally defined inflationary reheat temperature $T_{{\rm RH}}\simeq\sqrt{M_{\rm Pl}\Gamma_\Phi}$, where $\Gamma_\Phi$ is the inflaton decay rate. This observation can lead to some interesting departures from ``vanilla'' UV freeze-in in certain cases, most notably allowing for the production of states with mass: $T_{\rm RH}<m<T_{\rm max}$.

The value of $T_{\rm max}$ is determined by the initial energy density of the inflaton field $\Phi$ via $T_{\rm max} \simeq  \rho_{\Phi, I}^{1/4}$.  The (model dependent) choice taken in  \cite{Giudice:2000ex} is $ \rho_{\Phi,I}\sim M_\Phi^2M_{\rm Pl}^2$, then parameterising $\Gamma_\Phi\simeq \kappa^2M_\Phi$ one has that 
\beq
\frac{T_{\rm max}}{T_{\rm{RH}}}\simeq\frac{ \rho_{\Phi, I}^{1/4} }{\sqrt{M_{\rm Pl}\Gamma_\Phi}}
 \simeq \frac{1}{\kappa}~.
\eeq
Observe that the coupling  $\kappa$ in the decay rate introduces a degree of freedom to introduce a hierarchy between $T_{\rm max}$ and $T_{\rm RH}$. Other models of inflation lead to different forms for $\rho_{\Phi,I}$ and thus alter the ratio of $T_{\rm max}/T_{\rm RH}$; given this we treat the ratio $T_{\rm max}/T_{\rm RH}$ as a free parameter here. We will return to re-examine this connection between inflationary reheating and UV freeze-in in a forthcoming paper.

Importantly, at temperatures in excess of $T_{\rm RH}$, the inflaton field is continually decaying, and as such, entropy is not conserved in the thermal bath; this significantly alters the Boltzmann equations compared to those for $T<T_{\rm RH}$. As a result, the temperature of the thermal bath scales as $T \propto a^{-3/8}$ during the inflationary era, in contrast to the typical $T \propto a^{-1}$ during radiation domination.  Notably, it is precisely this regime that is of interest in the Boltzmann suppressed UV freeze-in scenario with $m_{\rm DM}>T_{\rm RH}$.

A careful analytic treatment of the Boltzmann equations in the regime $T>T_{\rm RH}$ is given in \cite{Giudice:2000ex}, which we very briefly sketch below. We can divide our analysis into three cases: 
\begin{itemize}
\item Relativistic production: $T_{\rm{RH}}<  m_{\rm DM}  \lesssim T_{\rm max}$.
\item Non-relativistic production: $ T_{\rm max} < m_{\rm DM} \lesssim 4T_{\rm max} $ . 
\item Exponentially suppressed production: $4T_{\rm max} \lesssim m_{\rm DM}$.
\end{itemize}
In the former case, we have non-relativistic production of dark matter particles (as the bath temperature is less than the dark matter mass at all times), and solving the Boltzmann equation gives the relic density as
\begin{equation} \label{eq:3.21}
    \Omega_{\rm DM} h^2 = \frac{3 \sqrt{10}}{2048 \pi^6} \left(\frac{17}{2e}\right)^{17/2} \frac{g_{\rm DM}^2}{g_*^{3/2}} \left(\frac{m_{\rm DM}}{\Lambda}\right)^{n+2} \frac{M_{\rm{Pl}} T_{\rm{RH}}^7}{T_0 m_{\rm DM}^7} \Omega_R h^2~,
\end{equation}
where $T_0$ is the present temperature of the universe and $\Omega_R$ is the ratio of the present radiation density to the critical density. Maximal dark matter production occurs at $T \simeq m_{\rm DM}/4$, which we take as the freeze-in temperature.

On the other hand, when $T_{\rm{RH}} < m_{\rm DM} < T_{\rm max}$, we have relativistic production of dark matter particles until $T \sim m_{\rm DM}$ and then dark matter production becomes non-relativistic. We first find the contribution to the number density $n^{\rm{rel}}_{\rm{DM}}$ of the dark matter particles from solving the Boltzmann equation in the relativistic regime $T \gtrsim m_{\rm DM}$
\begin{equation} \label{eq:3.22}
    n^{\rm{rel}}_{\rm DM} = \frac{1}{6-n} \frac{8}{\sqrt{5 \pi^{11}}} \frac{1}{\sqrt{g_*}} \left( \frac{m_{\rm DM}}{\Lambda} \right)^{n+2} M_{\rm{Pl}} T_{\rm{RH}}^2~.
\end{equation}
In the relativistic regime, the maximal dark matter production occurs at the lowest possible temperature, i.e.~$T \sim m_{\rm DM}$. Taking eq.~(\ref{eq:3.22}) as the FIMP number density at $T \sim m_{\rm DM}$, we solve the Boltzmann equation in the non-relativistic regime for $T \lesssim m_{\rm DM}$ to find the present relic density. This amounts to rescaling the number density in eq.~(\ref{eq:3.22}) to the present day, obtaining the relic density, and then simply adding that to the relic density in eq.~(\ref{eq:3.21}).

  Then, to obtain the relativistic contribution to the dark matter particle relic abundance, we multiply eq.~(\ref{eq:3.22}) by the scaling factor to obtain
  \beq
  \frac{a^3(T=m_{\rm DM})}{a_0^3} = \frac{a^3(T=m_{\rm DM})}{a_{\rm{RH}}^3} \times \frac{a_{\rm{RH}}^3}{a_0^3} = \frac{T^8_{\rm{RH}}}{m_{\rm DM}^8} \times \frac{T_0^3}{T^3_{\rm{RH}}}~.
  \label{eq:3.22b}
  \eeq

Thus, in the case that $ T_{\rm max} < m_{\rm DM} \lesssim 4T_{\rm max} $, the dark matter particle relic abundance arises as the sum of the non-relativistic contribution from eq.~(\ref{eq:3.22}) and the relativistic contribution coming from eq.~(\ref{eq:3.22b}), which gives the form.
\begin{equation}
\begin{split}\label{3.25}
\Omega_{\rm DM} h^2 
& = \left[ \frac{10}{6-n} \frac{8}{\sqrt{5 \pi^{15}}} \frac{1}{\sqrt{g_*}} +  \frac{3 \sqrt{10}}{2048 \pi^6} \left(\frac{17}{2e}\right)^{17/2} \frac{g_{\rm DM}^2}{g_*^{3/2}}\right]  \left(\frac{m_{\rm DM}}{\Lambda}\right)^{n+2} \frac{M_{\rm{Pl}} T_{\rm{RH}}^7}{T_0 m_{\rm DM}^7} \Omega_R h^2~,
\end{split}
\end{equation}
where we have made use of the fact that the radiation density at present is $\rho_R \approx \frac{3 \pi^2}{30} T_0^4$. The two terms in the square brackets are comparable for both dimension-5 ($n=0$) and dimension-6 ($n=2$) cases (roughly of the order of $10^{-3}$), so the contributions from relativistic and nonrelativistic production are typically similar for these cases.

Finally, for $m_{\rm DM} \gtrsim 4 T_{\rm max}$, FIMP production is in the deep non-relativistic regime and the relic abundance becomes exponentially suppressed \cite{Giudice:2000ex}, and we do not explore this possibility. Notably, the value of $T_{\rm max}$ does not explicitly enter into the parametric form of $\Omega_{\rm DM}$ (cf.~eq.~(\ref{eq:3.21}) \& eq.~(\ref{3.25})), rather its main influence is in determining the upper ranges of $m_{\rm DM}$ for which relativistic and non-relativistic production can occur in accordance with the itemized list given previously.

\begin{figure}[t] 
    \centerline{
    \includegraphics[scale=0.72]{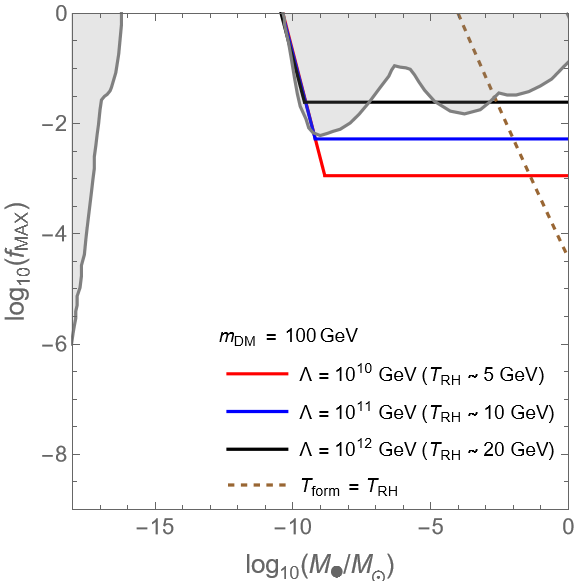} ~
    \includegraphics[scale=0.72]{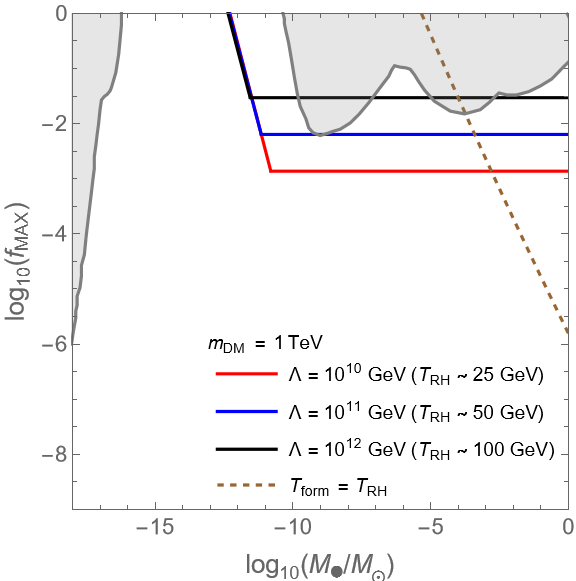} }
    \caption{{\bf Dimension 5.} \label{fig:dim-5}
    $f_{{\rm max}}$ vs $M_{\bullet}$ constraints for freeze-in freeze-in involving a dimension 5 operator $\frac{1}{\Lambda} \phi^{\dagger} \phi \bar{f} f$, where $\Lambda$ is the energy scale of new physics.  We consider the generic s-wave cross-section for $\phi \phi^{\dagger} \rightarrow b \bar{b}$ annihilations. We take three values of $\Lambda$ and the corresponding $T_{{\rm RH}}$ values are shown in parentheses. This is calculated in the Boltzmann suppressed UV freeze-in regime, where we assume $T_{\rm max} \sim 50 T_{{\rm RH}}$, for standard UV freeze-in with $m_{\rm DM}<T_{\rm RH}$, there is no constraint. The left panel is for 100 GeV dark matter, and the right panel is for 1 TeV dark matter. In both panels, FIMP equilibration with the Standard Model bath is avoided provided $\Lambda \gtrsim 10^{9}$ GeV.  PBH form prior to reheating in the parameter space left of the brown dashed line.
    }
\end{figure}

\begin{figure}[t] 
    \centerline{
    \includegraphics[scale=0.70]{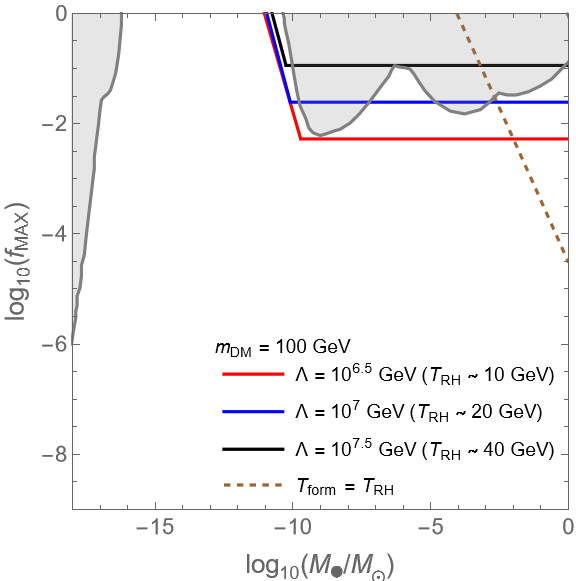} ~~~
    \includegraphics[scale=0.70]{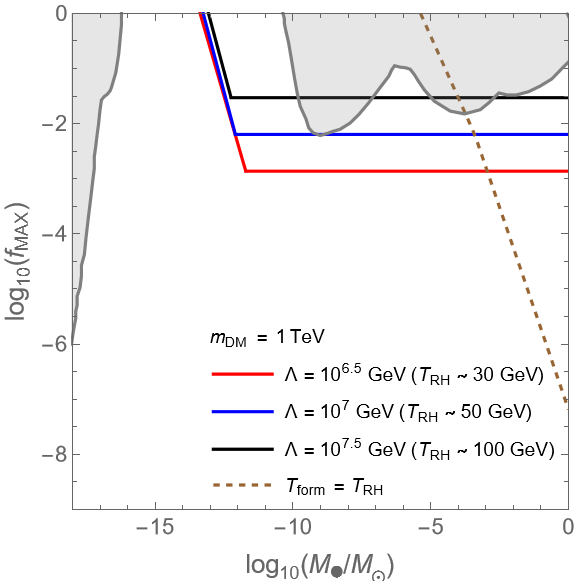} }
    \caption{{\bf Dimension 6.} \label{fig:dim6}
$f_{{\rm max}}$ vs $M_{\bullet}$ constraints for freeze-in involving a dimension 6 operator $\frac{1}{\Lambda^2} \bar{X} X \bar{f} f$. We consider the generic s-wave cross-section for $X \bar{X} \rightarrow b \bar{b}$ annihilations. Corresponding $T_{{\rm RH}}$ values to each value of $\Lambda$ are shown in parentheses. The left panel corresponds to 100 GeV dark matter, and the right panel corresponds to 1 TeV dark matter. We take three values of $\Lambda$ in both panels. As with dimension five for standard UV freeze-in with $m_{\rm DM}<T_{\rm RH}$ there are no constraints but for Boltzmann Suppressed UV freeze-in constraints can arise, here we take $T_{\rm max} \simeq 50 T_{{\rm RH}}$ as before.  In both panels, one can avoid FIMP equilibration with the Standard Model bath provided $\Lambda \gtrsim 10^{6}$ GeV.  PBH form prior to reheating in the parameter space left of the brown dashed line.}
\end{figure}

In Figures \ref{fig:dim-5} \& \ref{fig:dim6} we show the limits on $f_{\rm PBH}$ as a function of the PBH mass for the case of Boltzmann Suppressed UV freeze-in via a dimension five or dimension six portal operator, respectively. Recall that we found that vanilla UV freeze-in with $m_{\rm DM}<T_{\rm RH}$ dark matter annihilations did not constrain $f_{\rm PBH}$ to be less than unity. In contrast, for Boltzmann Suppressed UV freeze-in, limits can be obtained. Necessarily, we must specify the maximum temperature of the thermal bath (which is set by the details of inflationary reheating) and we take $T_{\rm max} \simeq 50 T_{{\rm RH}}$. In some regions of the parameter space that we present, we find that the PBH formation time is earlier than the point of reheating. We retain the exclusion curves in this region, but highlight that PBH formation with $t_{\rm form} < t_{\rm RH}$ requires an early matter-dominated period (or similar) prior to radiation-dominated. We mark the point $t_{\rm form} = t_{\rm RH}$ as a line on the plots appearing in Figures \ref{fig:dim-5} \& \ref{fig:dim6}.

\newpage

\clearpage

As noted above, the exact value of $T_{\rm max}$ is not important, since it only enters as a threshold compared to $m_{\rm DM}$ for determining whether relativistic production is possible, or indeed, if all production is exponentially suppressed.\footnote{This statement is true for operators of mass dimension lower than eight and assuming standard cosmology. For very high dimension operators and non-standard early universe cosmology then $T_{\rm max}$ will explicitly enter the expressions \cite{Bernal:2019mhf}, but we do not consider these cases here.} We confirm that the FIMPs do not come into equilibrium with the Standard Model bath, which places a lower limit on $\Lambda$ (see Appendix \ref{ApC}), and this restriction is noted in the caption.  Notably, for the case of Boltzmann Suppressed UV freeze-in, we obtain stronger limits on $f_{\rm PBH}$ than those coming from standard PBH limits (e.g.~microlensing).

\begin{figure}[t] 
    \centerline{
    \includegraphics[scale=0.7]{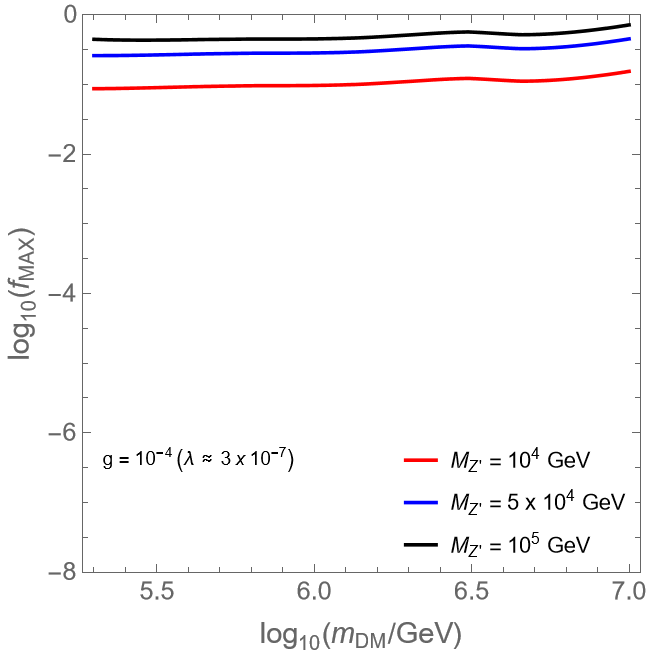} ~
    \includegraphics[scale=0.79]{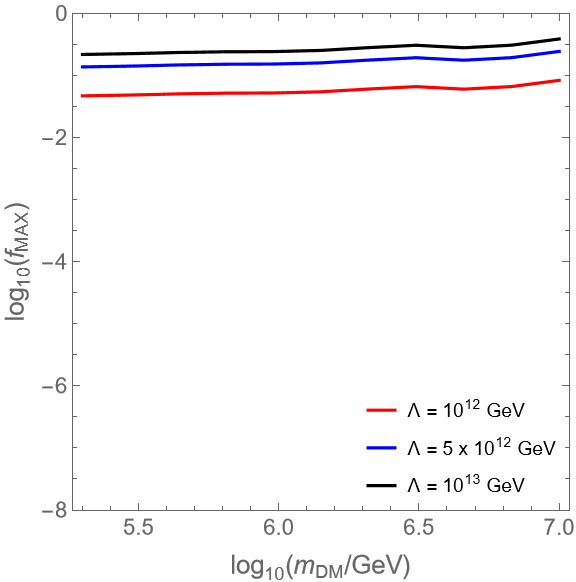} }
    \caption{{\bf Heavy FIMPs.}  {\bf Left Panel [Dimension 4].} We show the limits for the IR freeze-in model where the dark matter interacts with the Standard Model via a renormalizable interaction $Z'$. We consider three values of $M_{Z'}$. For the coupling values of $g$ and $\lambda$ used, the dominant freeze-in channel is $f \bar{f} \rightarrow X \bar{X}$ and the dominant annihilation channel is the t-channel cascade $X \bar{X} \rightarrow Z' Z' \rightarrow b \bar{b} b \bar{b}$. Lower $M_{Z'}$ leads to more stringent constraints on $f_{\rm PBH}$, however, sector equilibration also becomes more dangerous for lighter $Z'$. For the range of $Z'$ masses we use in our plots, we find that sector equilibration is avoided for $g \simeq 10^{-3}$ (cf.~Appendix \ref{ApC}).
    {\bf Right Panel [Dimension 5].} Limits for Boltzmann suppressed UV freeze-in of superheavy FIMP via the dimension 5 operator $\frac{1}{\Lambda} \phi^{\dagger} \phi \bar{f} f$.  The correct relic abundance is found by an appropriate choice of $T_{\rm RH}$ where we take $T_{\rm max}=50T_{\rm RH}$ and  $T_{\rm RH}<m_{\rm DM}$. The plot holds for $M_\bullet\gtrsim10^{-15}M_\odot$. Sector equilibration is avoided for $\Lambda \gtrsim 10^{12}$.}
\label{fig:shdm}
\end{figure}

\subsection{Superheavy dark matter}

It is well known that superheavy particle dark matter generically cannot be realised with the framework of freeze-out, since for masses in excess of 100 TeV the annihilation cross-section required to reduce a thermal abundance to the observed dark matter relic density violates unitarity \cite{Griest:1989wd}. In contrast, there is nothing that precludes the possibility of superheavy particle dark matter in the context of freeze-in (or FIMPZILLAs).  FIMPs can potentially have masses approaching the Planck scale, with observations of the CMB primordial tensor-to-scalar ratio constraining the mass to be  $m_{\rm DM}\lesssim 0.01M_{\rm Pl}$ with reasonable assumptions \cite{Garny:2015sjg}. Notably, annihilations of heavy FIMPs around PBHs can still lead to constraints and potentially observable signals.

Here, we examine the Boltzmann suppressed case for IR freeze-in via a $Z'$ (cf.~\cite{Belanger:2024yoj,Arcadi:2024obp,Boddy:2024vgt}) and UV freeze-in via a dimension 5 operator. To approach the study of superheavy FIMPs, we adopt a slightly different methodology from the previous sections.
We proceed by adapting existing limits on heavy decaying particle dark matter. One can convert the lower bound on dark matter lifetime $\tau_{\rm{DM}}$ to an upper bound on the PBH fraction via the matching \cite{Adamek:2019gns,Boucenna:2017ghj}
\begin{equation}\label{tau}
    f_{\rm PBH} = \frac{M_{\bullet} }{\Gamma_{\rm PBH}  \tau_{\rm DM} m_{\rm DM}}~.
\end{equation}  
We adapt the decaying dark matter limits of \cite{Skrzypek:2022hpy} based on Fermi-LAT observations of diffuse extragalactic $\gamma$-rays \cite{Fermi-LAT:2014ryh}. 
Taking the lower bounds on $\tau_{\rm{DM}}$ derived in \cite{Skrzypek:2022hpy} (using the central value) and reinterpreting this via eq.~(\ref{tau}), we obtain an upper bound on the fractional abundance of PBH $f_{\rm{max}}$ as a function of the particle dark matter mass $m_{\rm{DM}}$. We show the resulting limits  $f_{\rm PBH}<f_{\rm max}$ in Figure~\ref{fig:shdm} for particle dark matter masses in the range 100~TeV to 10~PeV.
As with previous plots, we check that the FIMP states do not come into equilibrium with the Standard Model thermal bath (the restriction on the coupling or mass scale $\Lambda$ is given in the caption). Notably, it can be seen in earlier figures that the value on $f_{\rm max}$ becomes independent of PBH mass for sufficiently heavy dark matter (assuming velocity-independent dark matter annihilations \cite{Chanda:2022hls}), and we also find $f_{\rm max}$ is largely independent of PBH mass in this analysis (for the UV freeze-in case, PBH mass dependence enters for very light PBH with $M_\bullet\lesssim10^{-15} M_\odot$). This is the reason why we break with the style of Figures \ref{fig:dim4}-\ref{fig:dim6} and plot $f_{\rm max}$ vs   $m_{\rm DM}$ in Figure \ref{fig:shdm} (rather than $f_{\rm max}$ vs PBH mass $M_{\bullet}$).

\section{Concluding Remarks}
\label{S4}

Many studies have examined the possibility that the dark matter relic density $\Omega_{\rm DM}$ could arise as a mixture of PBH and WIMPs. Here we have presented the first dedicated study to explore the implications for indirect detection for $\Omega_{\rm DM}$ arising as a mixture of PBH and non-WIMP dark matter. We have focused on the case of FIMP dark matter particles, examining renormalisable freeze-in, UV freeze-in, as well as Boltzmann suppressed freeze-in (with $T_{\rm RH}<m_{\rm DM}$ as outlined in \cite{Giudice:2000ex}, see also \cite{Cosme:2023xpa,Arcadi:2024obp,Bernal:2024ndy,Belanger:2024yoj,Boddy:2024vgt,Bernal:2025xx}).

Notably, we have seen through Figures  \ref{fig:dim4}-\ref{fig:shdm} that limits on the mixed PBH-FIMP scenarios can arise in a variety of freeze-in models. Importantly, these limits can be stronger than those arising solely from considering the observation implications of PBH, such as evaporations, gravitational waves, lensing, and CMB distortions (see e.g.~\cite{Carr:2020gox}). For certain FIMP models, the range of exclusions extends to fill the asteroid-mass PBH window (cf.~Figure~\ref{fig:dim4}), which is unconstrained by limits on PBH alone. Notably, while one does not expect indirect detection signals from FIMPs, the presence of PBHs can enhance these signals, making them potentially observable at future indirect detection experiments.

There are some interesting variants that we have not explored in this work. In particular,  it can be seen from  Figure \ref{fig:dim4} that the limits are stronger for lighter FIMPs. Thus, one might consider the case of sub-GeV FIMPs for which indirect detection limits based on annihilations to $b$-quarks (as utilized in this work) are inappropriate. For instance, particle dark matter with keV-scale mass would imply $X\bar{X}\rightarrow e^+e^-$ as the likely annihilation channel \cite{Chang:2019xva}. This is an interesting prospect, but dedicated analyses would be required to place limits on this mixed PBH-keV FIMP scenario. Moreover, while we have focused on the constraints coming from $\gamma$-ray fluxes, it would be interesting to explore the limits on mixed FIMP-PBH scenarios coming from other observables, such as radio waves or CMB measurements. Notably, such studies have previously been explored in the context of mixed WIMP-PBH scenarios \cite{Gines:2022qzy,Tashiro:2021xnj}, so it would be natural to extend these to the FIMP case.

Finally, it could be interesting to explore non-minimal variants. For instance, in exploring the constraints on models of Boltzmann suppressed UV freeze-in, in which freeze-in occurs during the process of inflationary reheating \cite{Giudice:2000ex}, we have assumed the inflationary dynamics of simple. However, it has been shown that UV freeze-in with non-standard cosmologies can lead to significant enhancements to the FIMP production rate \cite{Bernal:2019mhf,Bernal:2025fdr,Bernal:2024ndy,Bernal:2025xx}. Alternatively, such models in which the FIMP particles can annihilate to light hidden sector states, as in the ``Dark Sink'' scenario proposed in \cite{Bhattiprolu:2023akk}. Such a dark sink allows for larger couplings between the hidden and visible sector, and thus larger annihilation rates, and would also alter the FIMP halo profiles around the PBH.

\vspace{2mm}
 {\bf Acknowledgments.}
We thank Basudeb Dasgupta and, especially, Jakub Scholtz for insightful interactions. We are also grateful to the anonymous referee for their helpful remarks. JU is supported by NSF grant PHY-2209998.

\appendix

\section{PBH Dark Matter Halo Profiles}
\label{ApA}

As shown in Boudad {\em et al.} \cite{Boudaud:2021irr}, the particle dark matter halos around PBHs can be complicated when computed carefully. The analysis of \cite{Boudaud:2021irr} particle dark matter pair annihilations is absent. Pair annihilations were subsequently included in \cite{Chanda:2022hls}, highlighting that these processes significantly impact the late-time density profile in the case of WIMPs.  
The purpose of this Appendix is to summarise the results of Boudad {\em et al.} \cite{Boudaud:2021irr}. This is used as a starting point in Section \ref{sec2.2}, in which we will identify the dark matter particles as FIMPs and assess the impact of the FIMP pair annihilation rate on the PBH halo.

To study the mechanics of halo formation, let us first consider the equation describing the kinematics of a shell of particles at a radius $r$ from the PBH (see e.g.~\cite{Adamek:2019gns}):
\begin{equation}
    \ddot{r} = -\frac{G M_{\bullet}}{r^2} + \frac{\ddot{a}}{a}r~,
\end{equation}
where the first term on the right describes the effect of Newtonian gravitational attraction of the PBH, and the second term is the background deceleration (assuming radiation domination) of the FLRW spacetime. The time when these two terms become comparable is the turnaround time, which marks the time when the particles on the shell at radius $r$ decouple from the Hubble flow and re-collapse towards the black hole.

One can define a quantity called the `turnaround radius', defined as   \cite{Adamek:2019gns} 
\beq \label{rta}
r_{{\rm ta}}(t)\simeq (2 G M_{\bullet}  t^2)^{1/3}~,
\eeq
 at some cosmic time $t$ during radiation domination, which can be related to temperature via
 \beq
 \label{tT}
 t = \frac{1}{T^2}\sqrt{\frac{45}{16 \pi^3 G g_*}}~.
 \eeq  
 This turnaround radius is the radius at which the shell of particles gets decoupled from the Hubble flow and starts falling towards the PBH at time $t$.

The approximate density distribution in the halo around the PBH can be expressed as a function of the normalized (in units of the Schwarzschild radius $r_s=2GM_{\bullet}$) distance $\tilde{r} = r/r_s$ as follows  \cite{Boudaud:2021irr}
\begin{equation} \label{eq:rhor}
    \rho(\tilde{r}) = \sqrt{\frac{2}{\pi^3}} \frac{\rho_i}{\sigma^3_i} \tilde{r}^{-3/2} \int \int {\rm d} \mathcal{R} {\rm d}u \left\{ \mathcal{R}(1-u) \right\}^{3/2} \Theta (\bar{u}(\tilde{r})-u) \int_{\sqrt{\mathcal{Y}_m} \Theta(\mathcal{Y}_m)}^1 \frac{{\rm d}y}{\sqrt{y^2 - \mathcal{Y}_m}}~,
\end{equation}
with $\bar{u}(\tilde{r})=\tilde{r} \sigma_{\rm collapse}^2(\tilde{r}) $ and $\mathcal{Y}_m = 1 + \mathcal{R}^2 \left\{ \frac{1}{u} \left(1 - \frac{1}{\mathcal{R}}\right) - 1 \right\}$.

The density profile of particle dark matter around a PBH is calculated by evaluating the integral in eq.~(\ref{eq:rhor}), which has three distinct analytic forms corresponding to
\begin{itemize}
 \item[(i).] $\bar{u}(\tilde{r}_{\rm eq}) < \bar{u}(\tilde{r}_{i}) < 1$, 
 \item[(ii).] $\bar{u}(\tilde{r}_{\rm eq}) < 1 < \bar{u}(\tilde{r}_{i})$, 
 \item[(iii).] $1 < \bar{u}(\tilde{r}_{\rm eq}) < \bar{u}(\tilde{r}_{i})$.
 \end{itemize}
 These conditions can be recast in terms of the mass of the PBH relative to the following two mass scales 
\begin{equation}
    M_1 = \left( \frac{T'_i}{m_{\rm DM}} \right)^{3/2} \frac{\eta_{\rm ta}^{1/2} t_i }{2G}~,
\end{equation}
and 
\begin{equation}
    M_2 =  \left( \frac{T'_i}{m_{\rm DM}} \right)^{3/2} \frac{\eta_{\rm ta}^{1/2} t_{\rm eq} }{2G} \left( \frac{T_{\rm eq}}{T_i} \right)^3~.
\end{equation}
Thus, following \cite{Boudaud:2021irr}, one can classify the three regimes into a heavy PBH case, an intermediate PBH case, and a light PBH case. Each of the three regimes corresponds to a different set of nested power laws, as we summarise below. We also note that these forms give the profile of the halo prior to considering dark matter interactions or stripping due to close encounters.   The reader is directed to  \cite{Boudaud:2021irr,Chanda:2022hls} for more detailed discussions and derivations.

\subsection{Heavy PBH regime}
For heavier PBH for which  $M_{\bullet} > M_1 > M_2$ then the density profile has the following form
\begin{equation}
    \rho(\tilde{r}) = \begin{cases}
        \rho_{3/2}(\tilde{r}) &~~~~~ 0<\tilde{r}<\tilde{r}_1, \\
        \rho_{9/4}(\tilde{r}) &~~~~~ \tilde{r}_1<\tilde{r}<\tilde{r}_{\rm eq}, \\
        0 &~~~~~ \tilde{r}_{\rm eq}<\tilde{r},
    \end{cases}
\end{equation}
where 
\begin{equation}
    \rho_{3/2}(\tilde{r})=\sqrt{\frac{2}{\pi^3}} \rho_{i}\tilde{r}^{3/2}_{i}\left[ \frac{2 \sqrt{2 \pi}}{3} \left\{2 + \left(1+2\frac{\tilde{r}}{\tilde{r}_{i}}\right) \sqrt{1-\frac{\tilde{r}}{\tilde{r}_{i}}}\right\}\right]\tilde{r}^{-3/2}~,
\end{equation}
and
\begin{equation}
    \rho_{9/4}(\tilde{r})= \frac{\sqrt{128 \pi}}{\Gamma^2(4)} \rho_{\rm eq}\tilde{r}_{\rm eq}^{9/4} \left\{1-\frac{\Gamma^2(4)}{3\sqrt{2\pi^3}}\left(\frac{\tilde{r}}{\tilde{r}_{\rm eq}}\right)^{3/4}\right\}\tilde{r}^{-9/4}~,
\end{equation}
where we have defined $\rho_{\rm eq} = \rho_{\rm collapse}(\tilde{r}_{\rm eq})$. The transition point $\tilde{r}_1$ is determined by demanding continuity of $\rho(\tilde{r})$ at $\tilde{r}=\tilde{r}_1$.

\subsection{Intermediate PBH regime}
If the mass of the PBH is such that it falls in the intermediate regime $M_1 > M_{\bullet} > M_2$, then the halo of particle dark matter will follow the nested power-laws below
\begin{equation}
    \rho(\tilde{r}) = \begin{cases}
        \rho_{3/4}(\tilde{r}) &~~~~~ 0<\tilde{r}<\tilde{r}_1 \\
        \rho'_{3/2}(\tilde{r}) &~~~~~ \tilde{r}_1<\tilde{r}<\tilde{r}_2 \\
        \rho_{9/4}(\tilde{r}) &~~~~~ \tilde{r}_2<\tilde{r}<\tilde{r}_{\rm eq} \\
        0 &~~~~~ \tilde{r}_{\rm eq}<\tilde{r}
    \end{cases}~.
\end{equation}
The innermost density profile is $\rho\propto\tilde{r}^{-3/4}$; the full expression is below
\begin{equation}
    \rho_{3/4}(\tilde{r}) \simeq 4.2 \sqrt{\frac{2^{5/2}}{\pi^3}}  \Gamma\left(\frac{7}{4}\right) \frac{\rho_{i}}{\sigma_{i}^{3/2}}\tilde{r}^{-3/4}~.
\end{equation}
The outermost profile is $\rho\propto \tilde{r}^{-9/4}$ (and is identical to the heavy PBH case). There is also a transition between these two regions in which $\rho\propto \tilde{r}^{-3/2}$, but where the form is different from the heavy PBH case (denoted by the dash), and in this region the density is given by
\begin{equation}
    \rho'_{3/2}(\tilde{r}) = \sqrt{\frac{2}{\pi^3}} \frac{\rho_{i}}{\sigma^3_{i}} \left(1.047-\frac{3 \pi}{8} \frac{\tilde{r}}{\tilde{r}_{\rm eq}} \right) \tilde{r}^{-3/2}~,
\end{equation}
The transition points $\tilde{r}_1$ and $\tilde{r}_2$ can be found as before by matching the corresponding profiles at those points and then solving them.

\subsection{Light PBH regime }
Finally, there is the light PBH case for which  $M_1 > M_2 > M_{\bullet}$. In this case, the density profile is
\begin{equation}
    \rho(\tilde{r}) = \begin{cases}
        \rho_{3/4}(\tilde{r}) &~~~~~ 0<\tilde{r}<\tilde{r}_1 \\
        \rho'_{3/2}(\tilde{r}) &~~~~~ \tilde{r}_1<\tilde{r}<\tilde{r}_{\rm eq} \\
        0 &~~~~~ \tilde{r}_{\rm eq}<\tilde{r}
    \end{cases}~.
\end{equation}
The expressions for $\rho_{3/4}$ and $\rho'_{3/2}$ are the same as in the intermediate PBH regime. Again, $\tilde{r}_1$ can be determined by matching these profiles.

\section{Neglecting Early Abundances of Freeze-in Dark Matter Particles}
\label{ApB}

In this appendix, we examine whether it is reasonable to assume that the density of dark matter particles around the PBH is uniform at $T=T_{\rm FI}$ for the case of IR freeze-in. As noted in Section \ref{S2.1}, IR freeze-in is complicated as the FIMPs are gradually populated over an extended period. Here we shall argue that it is reasonable to assume that the density of FIMPs around the PBH is uniform at $T=T_{\rm FI}$ since 
\begin{itemize}
\item[ i).]~The vast majority of FIMPs are produced at a characteristic time scale $t_{\rm FI}$;
\item[ii).]~The infall time for dark matter is longer than the characteristic time scale for freeze-in. 
\end{itemize}

\subsection{Early FIMP abundance}

The FIMP abundance due to IR freeze-in via a $Z'$ was calculated in Section \ref{4.1}. Specifically, let us consider the scenario with $M_{Z'} < m_{\rm DM} $.
 We will take the matrix element to be that of eq.~(\ref{MM}) of Case A.i. It follows from the Boltzmann equation that between two temperature values  (with $T\equiv m/x$) an abundance of FIMPs is generated according to
\beq\label{dY}
\Delta Y=  \frac{45}{256 \pi^7} \frac{g^2\lambda^2 M_{\rm Pl}}{1.66 \cdot g_*^S\sqrt{g_*^\rho}m_{\rm DM}}
\mathcal{I}(x_{i},x_f) \quad{\rm with}\quad \mathcal{I}(x_{i},x_f)=\int_{x_i}^{x_f}  \big(x K_1(x)\big)^2~{\rm d}x~.
\eeq
In IR freeze-in, the period of maximal production is around $M_{Z'}$, however, we should quantify to what degree this late-produced FIMP abundance dominates the early abundance. To examine this, we will break the integral $\mathcal{I}(x_{i},x_f)$  into three parts: an early part with $0<x<x_e$, a late-time production period $x_l<x<\infty)$, and the peak production $x_e<x<x_l$. We anticipate that $x_e,x_l\sim 1$. We can identify the numerical values for $x_e$ and $x_l$ by identifying when the values that lead to the early and late periods become subdominant.

Firstly, one can verify numerically that for $x_e\approx0.54$ one has  $\mathcal{I}(0,x_e)\approx\mathcal{I}(x_e,\infty)$. Moreover, for $x_l\approx1.42$ one has that $\mathcal{I}(x_e,x_l)\approx\mathcal{I}(x_l,\infty)$. This is a model-independent statement that only depends on the properties of the mathematical function defining $\mathcal{I}$. Note that $x_e$ compares to $T\simeq 2m_{\rm DM}$ and  $x_l$ compares to $T\simeq 0.7m_{\rm DM}$. Thus a relatively small temperature range $m_{\rm DM}/4 <T< 4m_{\rm DM}$  accounts for 75\% of the total abundance. Thus, only a small fraction of the dark matter is produced earlier (25\%) or later (0.1\%)  than this peak period for which $T\sim m_{\rm DM}$. Hence, at time $t\sim t_{\rm FI}$ we can reasonably consider most of the FIMPs to be freshly produced and will form a uniform density distribution around the PBH since it is sourced from interactions from bath particles which are free-streaming and not bound by the gravitational potential of the PBH. Moreover, if any earlier population of FIMPs has started to undergo collapse, it is a negligible perturbation away from the assumption of a uniform density distribution of particle dark matter at time $t\sim t_{\rm FI}$. This is especially true given that any density increase of FIMP dark matter at the centre will be dwarfed by the central PBH.

\subsection{Collapse time}

We can use the standard time-temperature relationship for a radiation-dominated universe, as given in eq.~(\ref{tT}), 
to estimate the time of freeze-in assuming renomralisable interactions 
\beq
t_{\rm FI, IR}
\sim \frac{M_{\rm Pl}}{\left({\rm max}[m_{\rm DM},M_{Z'}]\right)^2}
\sim 10^{-12}~{\rm s}\left(\frac{500~{\rm GeV}}{{\rm max}[m_{\rm DM},M_{Z'}]}\right)^2~.
\eeq
Next, we estimate the collapse time. 
Since we have assumed that the particle dark matter has no self-interactions, the evolution of the density profile is purely gravitational. Moreover, we anticipate from an earlier study \cite{Chanda:2022hls} that the total mass of the late-time halo will be comparable to the PBH mass. Thus, we expect that (especially for earlier halos) the PBH mass largely determines the gravitational potential. Accordingly, the collapse time for the dark matter around the PBH is characteristically the free-fall time of an individual particle, given by the Newtonian relationship
\beq\label{eq:t_ff}
t_{\rm ff}\simeq \sqrt{\frac{r^3}{2GM_{\bullet}}} ,
\eeq
where $r$ is the distance of a dark matter particle from the PBH. To estimate how long it takes for the gravitational potential to be significantly perturbed, let us calculate the free-fall time for a particle with initial position $r=r_{\rm eq}/2$. Recall that $r_{\rm eq}$ is the turnaround radius (cf.~eq.~(\ref{rta}) at matter-radiation equality $t_{\rm eq}\approx 2.4\times10^{12}$s  demarks the extent of the PBH halo, this is given by \cite{Adamek:2019gns}  
\beq
r_{\rm eq}\simeq  \left(2GM_{\bullet}t_{\rm eq}^{2}\right)^{1/3}~.
\eeq 
We can calculate the free fall time for a particle at a distance $r_{\rm eq}/2$ is given by 
\beq\label{eq:t_ff2}
t_{\rm ff}\big|_{r=r_{\rm eq}/2}\simeq 
\sqrt{\frac{GM_{\bullet}t_{\rm eq}^{2}}{8GM_{\bullet}}} \simeq \frac{t_{\rm eq}}{8} \sim 10^{11}s~. 
\eeq
Comparing $t_{\rm ff}$ to $t_{\rm FI}$ we see that the temporal separation is vast, thus we do not expect the subpopulation of early produced FIMPs in the gravitational potential of PBH to undergo significant collapse prior to the completion of freeze-in at $t\sim t_{\rm FI}$.

\section{Computing Limits on the Maximal Fractional PBH Abundance}
\label{ApD}

In this Appendix, we give further details on our method of computing limits from Fermi-LAT data, and outline deviations from the analysis method employed in \cite{Chanda:2022hls}.
Here to obtain a bound on $f_{\rm{max}}$, the maximal fractional PBH abundance, we define a distance function
\begin{equation}
\Delta_\Phi(E_{\gamma}, f_{\rm{PBH}}) = {\rm log}\left( \frac{{\rm d} \Phi}{{\rm d}E}\right)\Big|_{f_{\rm{PBH}} } - {\rm log} \left( \frac{{\rm d} \Phi}{{\rm d} E} \right) \Big|_{\rm{Fermi-LAT}}~.
\end{equation}
This is a measure of how far (logarithmically) the calculated flux is from the Fermi-LAT flux, as a function of the photon energy $E_{\gamma}$ and $f_{\rm{PBH}}$. Note that the calculated flux $\frac{{\rm d}\Phi}{{\rm d}E}|_{f_{\rm{PBH}}}$ depends linearly on the fractional PBH abundance $f_{\rm{PBH}}$ as
\begin{equation}
    \frac{{\rm d} \Phi}{{\rm d} E}\Big|_{f_{\rm{PBH}} = f} = f \times \frac{{\rm d} \Phi}{{\rm d} E} \Big|_{f_{\rm{PBH}}  = 1}~.
\end{equation}
Then, at $f_{\rm{PBH}} = f$, we have 
\begin{equation}
\begin{split}
\Delta_\Phi(E_{\gamma}, f_{\rm{PBH}} = f)  &= {\rm log}\left( \frac{{\rm d} \Phi}{{\rm d}E}\right) \Big|_{f_{\rm{PBH}} = 1} - {\rm log} \left( \frac{{\rm d} \Phi}{{\rm d} E} \right) \Big|_{\rm{Fermi-LAT}} + {\rm log}(f) \\
&= \Delta_\Phi(E_{\gamma}, f_{\rm{PBH}}=1) + {\rm log}(f)~.
\end{split}
\end{equation}
For a given value $f_{\rm{PBH}} = f$, we need to ensure that the calculated flux is below the observed flux for all energies $E_{\gamma}$. Thus, if the maxima of $\Delta_\Phi(E_{\gamma}, f_{\rm{PBH}} = f) $ is at $E_{\gamma} = E_{\max}$, we need 
\begin{equation} \label{eq:ii}
    \Delta_{\Phi}(E_\gamma = E_{\rm{max}}, f_{\rm{PBH}} = f) =   \Delta_\Phi(E_{\gamma} = E_{\rm{max}}, f_{\rm{PBH}}=1) + {\rm log}(f)\leq 0~.
    \end{equation}
 $f_{\rm{max}}$  is then precisely that value of $f$ for which equality is satisfied. Hence we obtain $f_{\rm{max}}$ by the requirement
 \begin{equation}
     {\rm log} (f_{\rm{max}}) = -\Delta_\Phi(E_{\gamma} = E_{{\rm max}}, f_{\rm{PBH}}=1)~.
 \end{equation}

\begin{figure}[t] 
    \centerline{
    \includegraphics[scale=0.7]{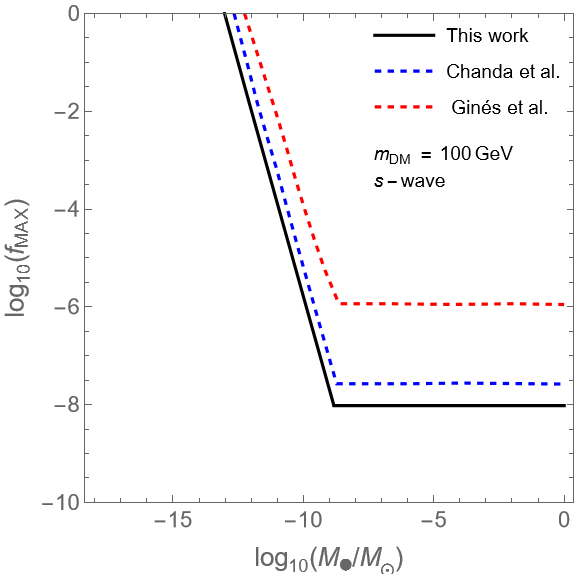} ~\hspace{15mm}
    \includegraphics[scale=0.7]{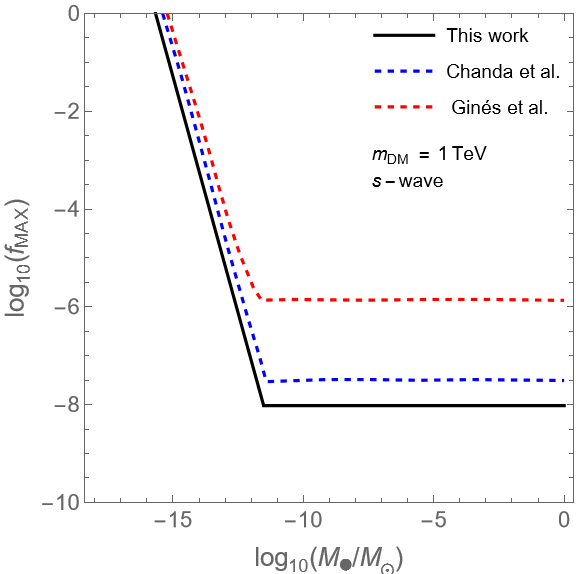} }
    \caption{We apply our analysis set-up to the case of s-wave freeze-out dark matter (as opposed to freeze-in) to obtain a comparison of the resulting limits to those of earlier studies  \cite{Chanda:2022hls,Gines:2022qzy}.  }
\label{fig:comp}
\end{figure}

This differs from the analysis in \cite{Chanda:2022hls}, wherein the authors extracted bounds $f_{\rm max}$ by identifying the value of $f_{\rm PBH}$ for which the peak of the differential flux exceeded the background observed by Fermi-LAT. In Figure \ref{fig:comp} we provide a comparison of the resulting limits which arise from using the method above to bounds derived in earlier papers  \cite{Chanda:2022hls,Gines:2022qzy}. In order to make the comparison, we apply the limits to the case of s-wave freeze-out dark matter (rather than freeze-in as assumed in the main text). We find that these slight changes in our comparison to data lead to an $\mathcal{O}(1)$ change in the limit to an earlier analysis by a subset of authors \cite{Chanda:2022hls}. Similar to \cite{Chanda:2022hls}, there remains a disconnect with the findings of \cite{Gines:2022qzy}.

\section{Sector Thermalization}
\label{ApC}

If the visible sector and FIMP sector come into thermal equilibrium, this spoils the freeze-in mechanism. After sector equilibration, the dark matter particles will subsequently freeze out, leading to a relic density that is determined by the freeze-out dynamics and insensitive to the earlier cosmological history. While this provides a consistent cosmological history, it is not the present case of interest.  
To avoid sector equilibration from occurring, one must ensure that the sectors are sufficiently decoupled. 
Specifically, we require that $n_{{\rm DM}} (T_{\rm FI}) \ll n_{{\rm eq}} (T_{\rm FI})$ and that the freeze-out temperature $T_{\rm{FO}} \gg T_{\rm{FI}}$.
The below provides a better criterion than those set out in  \cite{Elahi:2014fsa} (similar analyses appear elsewhere in the literature).

\subsection{Sub-thermal freeze-in abundance}

First, we consider the criteria $n_{{\rm DM}} (T_{\rm FI}) \ll n_{{\rm eq}} (T_{\rm FI})$, repeating a general arguement from   \cite{Elahi:2014fsa}. Suppose that UV freeze-in generates a near-thermal abundance at $T_{\rm RH}$ which evolves to the present day only due to changes in the entropy density of the universe, then 
\beq\label{ent}
 n_{{\rm DM}} (T_{\rm FI}) \simeq  n_{{\rm DM}} (T_0) \left[\frac{s(T_{\rm FI})}{s_0}\right]\simeq
\frac{ \rho_c \Omega_{\rm DM}}{m_{\rm DM}}\left[ \frac{g(T_{{\rm FI}})}{g(T_0)}\left(\frac{T_{{\rm FI}}}{T_0}\right)^3 \right]~,
\eeq 
where the ``0'' subscript indicates the values today. Requiring  $n_{\rm DM} (T_{\rm {FI}}) \ll T_{\rm {FI}}^3/\pi^2$ leads to a lower mass bound on the dark matter mass \cite{Elahi:2014fsa} 
\begin{equation}\label{condc}
m_{\rm DM} \gtrsim \frac{g(T_{{\rm FI}})}{g(T_0)} \frac{\pi^2 \rho_c \Omega_{\rm DM}}{1.2 T_0^3} \simeq 0.4 {\rm ~ keV}~.
\end{equation}
This bound holds for any typical (IR or UV)  operator that gives the observed relic density via freeze-in, provided there are no additional dark matter number-changing interactions. We next look at the decoupling requirements for UV freeze-in and then for a specific IR case.

\subsection{UV freeze-in}

For ``vanilla'' UV freeze-in, we have $m_{\rm DM} < T_{\rm{RH}}$ and the freeze-in temperature $T_{\rm{FI}} \sim T_{\rm{RH}}$. The condition that $n_{{\rm DM}} (T_{\rm RH}) \ll n_{{\rm eq}} (T_{\rm RH})$ implies \cite{Giudice:2000ex}: 
\beq\label{11111}
\Lambda \gg \left( \frac{8 M_{\rm{Pl}} T_{\rm{RH}}^{n+1}}{\sqrt{5} (6-n) \pi^{7/2} \sqrt{g_{* \rho}(T_{\rm{RH}})}} \right)^{1/(n+2)}.
\eeq
where $n=0$ corresponds to a dimension-5 and $n=2$ corresponds to a dimension-6 operator. Specifically,
\beq
\Lambda_5 &\gg 10^9 \rm{~GeV}  \sqrt{\frac{T_{\rm{RH}}}{100 \rm{~GeV}}}~, \\
 \Lambda_6 &\gg 10^7 \rm{~GeV}  \left( \frac{T_{\rm{RH}}}{10 \rm{~ TeV}} \right)^{3/4}~.
\eeq
Next, the condition  $T_{\rm FO}> T_{\rm RH}$ can be expressed as 
\begin{equation}
T_{{\rm RH}}\ll T_{{\rm FO}} \simeq \frac{1.66 \sqrt{g_*^{\rho}} \pi^2}{M_{\rm Pl} \langle \sigma v \rangle}.
\end{equation}
We have $\langle \sigma v \rangle \sim m_{\rm DM}^{n}/ \Lambda^{n+2}$ and then the above can be written on as a condition on the UV scale (assuming $m_{\rm DM}<T_{\rm RH}$ as standard for UV freeze-in) as
\beq \label{eq:C.6}
 \Lambda \gg \left(\frac{m_{\rm DM}^{n}  T_{{\rm RH}}M_{\rm Pl}}{1.66 \sqrt{g_*^{\rho}} \pi^2}\right)^{1/(n+2)}.
\eeq
Thus for dimension 5 $\langle \sigma v \rangle_{\phi \phi^{\dagger} \rightarrow f \bar{f}}\sim \Lambda^{-2}$ or dimension-6 operator $\langle \sigma v \rangle_{X \bar{X} \rightarrow f \bar{f}}\sim m_{\rm DM}^2/\Lambda^{4}$ one has
\beq\label{Lambda_5}
 \Lambda_5 &\gg 3 \times 10^9 \rm{~GeV}  \sqrt{\frac{T_{\rm{RH}}}{100 \rm{~GeV}}}~,\\
  \Lambda_6 & \gg 10^6 \rm{~GeV}  \left( \frac{T_{\rm{RH}}}{10 \rm{~ TeV}} \right)^{3/4}  \left( \frac{m_{\rm DM}}{100 \rm{~GeV}} \right)^{1/2}.
\eeq

\subsection{Boltzmann suppressed UV freeze-in}

In the Boltzmann suppressed UV-freeze in case, with $m_{\rm DM} > T_{\rm{RH}}$, again, we require $n_{{\rm DM}} (T_{\rm_{\rm FI}}) \ll n_{{\rm eq}} (T_{\rm FI})$ and this condition implies a restriction of the effective field theory scale \cite{Giudice:2000ex} 
\begin{equation}
\Lambda \gg \left( \frac{289 M_{\rm{Pl}} T_{\rm{RH}}^2 m_{\rm DM}^{n-1}}{2 \sqrt{10} \pi^{5/2} e^{17/4} \sqrt{g_*}} \right)^{\frac{1}{n+2}}.
\end{equation}
Again, we can find the limits for dimension-5 and dimension-6 cases as:
\beq
    \Lambda_5 &\gg 2 \times 10^9 \rm{~GeV}  \left(\frac{T_{\rm{RH}}}{100 \rm{~GeV}}\right)  \sqrt{\frac{100 \rm{~GeV}}{m_{\rm DM}}}~,\\ 
    \Lambda_6 &\gg 10^5 \rm{~GeV}  \sqrt{\frac{T_{\rm{RH}}}{10 \rm{~GeV}}}  \left( \frac{m_{\rm DM}}{100 \rm{~GeV}} \right)^{1/4}.
\eeq
For the condition $T_{\rm{FO}} \gg T_{\rm FI}$, we have $T_{\rm{FI}} \simeq m_{\rm DM}/4$ \cite{Giudice:2000ex}, and one can replace $T_{\rm{RH}} \rightarrow m_{\rm DM}/4$ in eq.~(\ref{eq:C.6}) and eq.~(\ref{Lambda_5}) to find the constraints on $\Lambda$ for the Boltzmann suppressed UV freeze-in case.

\subsection{IR freeze-in via $Z'$}
Let us  consider the case of IR freeze-in via a $Z'$ for the case when the freeze-in channel is via 2 $\rightarrow$2 scattering. We require,
\beq\label{Yeq}
Y_{\rm DM}\Big|_{T_{\rm{FI}}}\ll \frac{n_{\rm{eq}}}{s}\Big|_{T_{\rm{FI}}} \approx \frac{45}{\pi^4} \zeta(3) \frac{g_{\rm DM}}{g_{*S}}.
\eeq
For $M_{Z'}<2 m_{\rm DM}$ (referred to as {\bf Case A.i} in Section \ref{4.1}), $T_{\rm{FI}} \simeq m_{\rm{DM}}$ the yield is given by eq.~(\ref{eq:3.6}), and it follows:
\begin{equation}
    g \lambda \ll \sqrt{\frac{2^{14} \zeta(3) \pi \sqrt{g_{* \rho}} (1.66) m_{\rm DM}}{3 M_{\rm{Pl}}}} \sim 3 \times 10^{-6}  \sqrt{\frac{m_{\rm DM}}{300 \rm{~ GeV}}}~.
\end{equation}
Since $g \lambda \sim 3 \times 10^{-11}$ is determined by eq.~(\ref{eq:3.7}), this gives us a lower bound on $m_{\rm DM} \gg 30 \rm{~ eV}$.

On the other hand for $2 m_{\rm DM} <  M_{Z'}$  which we labelled {\bf Case A.ii}, taking $T_{\rm{FI}} \simeq M_{Z'}$ and using the yield from eq.~({\ref{eq:3.9}}) we find that avoiding equilibration requires
\beq
\left( \frac{1}{\lambda^2} + \frac{1}{g^2} \right) \gg \frac{405 M_{\rm{Pl}}}{512 \times 90 \times (1.66) \sqrt{g_{* \rho}} M_{Z'} \times \zeta(3)}~.
\eeq
Additionally, if we assume $g \ll \lambda$, then we have the constraint
\beq
g \ll \sqrt{\frac{512 \times 90 \times (1.66) \sqrt{g_{* \rho}} M_{Z'} \times \zeta(3)}{405 M_{\rm{Pl}}}}  \simeq  10^{-7}  \sqrt{\frac{M_{Z'}}{100 \rm{~GeV}}}~.
\eeq
The two-step freeze-in process,  referred to as {\bf Case B} in Section \ref{4.1}, is the relevant production process if $g \gg \lambda$ and $2 m_{\rm{DM}} < M_{Z'}$, in this case the yield is given by   (see e.g.~\cite{Scholtz:2019csj})
\beq
Y_{\rm{DM}} \simeq \frac{270}{64 \pi^4 g_{*S} \sqrt{g_{*\rho}}} \left( \frac{M_{\rm{Pl}} \lambda^2}{M_{Z'}} \right)~.
\eeq 
The corresponding equilibration constraint on $\lambda$ is 
\begin{equation}
\lambda \ll \frac{64 \zeta(3)}{3} \sqrt{g_{* \rho}} \left( \frac{M_{Z'}}{M_{\rm{Pl}}} \right) \simeq 5 \times 10^{-8}  \sqrt{\frac{M_{Z'}}{100 \rm{~ GeV}}}~.
\end{equation}

Next, we require $T_{\rm{FO}} \sim \frac{1.66 \sqrt{g_{* \rho}} \pi^2}{M_{\rm{Pl}} \langle \sigma v \rangle} \gg T_{\rm FI}$. For $m_{\rm DM} > M_{Z'}$, let's assume that $g \gg \lambda$ so that the primary annihilation channel is the t-channel process  $X\bar{X} \rightarrow Z'Z' $ (and thus sets the annihilation rate $\Gamma$). Specifically, if we take  the limit  $M_{Z'} \ll m_{\rm DM}$, the cross-section of eq.~(\ref{eq:4.10}) can be approximated as $\langle \sigma v \rangle_{X\bar{X} \rightarrow Z'Z'} \approx \frac{g^4}{4 \pi M_{Z'}^2}$. Also, $T_{\rm{FI}} \sim m_{\rm DM}$, so we have
\begin{equation}
    g \ll \left( \frac{1.66 \sqrt{g_{* \rho}} 4 \pi^3 M_{Z'}^2}{m_{\rm DM} M_{\rm{Pl}}} \right)^{1/4} \simeq 10^{-4}  \left( \frac{M_{Z'}}{300 \rm{~GeV}} \right)^{1/2}  \left( \frac{1 \rm{~TeV}}{m_{\rm DM}} \right)^{1/4}~.
\end{equation}

For the converse case $M_{Z'} > m_{\rm DM}$, the only annihilation channel possible is the s-channel process $X \bar{X} \rightarrow f \bar{f}$, and assuming $M_{Z'} \gg m_{\rm DM}$, we approximate eq.~(\ref{eq:3.13}) as $\langle \sigma v \rangle_{X \bar{X} \rightarrow f \bar{f}} \approx \frac{6 \lambda^2 g^2 m_{\rm DM}^2}{\pi M_{Z'}^4}$. Also, $T_{\rm{FI}} \sim M_{Z'}$, and thus we have
\begin{equation}
    \lambda g \ll \left( \frac{1.66 \sqrt{g_{* \rho}} \pi^3 M_{Z'}^3}{6 M_{\rm{Pl}} m_{\rm DM}^2} \right)^{1/2} \approx 10^{-8}  \left( \frac{M_{Z'}}{100 \rm{~GeV}} \right)^{3/2}  \left(\frac{100 \rm{~ GeV}}{m_{\rm DM}}\right)~.
\end{equation}
We see that for the specific FIMP models that we have examined here, requiring $T_{\rm{FO}} \gg T_{\rm FI}$ satisfying eq.~(\ref{Yeq}) tends to simultaneously satisfy $T_{\rm{FO}} \gg T_{\rm FI}$. However, this is not always the case since the two restrictions have different parameter dependencies.

   \end{document}